\documentclass[12pt, a4paper]{article}
\usepackage[font=footnotesize]{caption}
\usepackage[utf8]{inputenc}
\usepackage{graphicx}
\usepackage{authblk}
\usepackage{booktabs}
\usepackage{verbatim}
\usepackage{url}
\usepackage{caption}
\usepackage{subcaption}
\usepackage{indentfirst}
\usepackage{url}
\usepackage{amssymb}
\usepackage{amsmath}

\usepackage{todonotes}

\usepackage{rotating}
\usepackage{array}
\title{A Systematic Approach for Studying How Topological Measurements Respond to Complex Networks Modifications}
\author{Alexandre Benatti$^1$, Roberto M. Cesar Jr.$^1$, and \\ Luciano da F. Costa$^2$}

\affil{
$^1$Institute of Mathematics and Statistics - DCC \\
University of S\~ao Paulo \\
Rua do Mat\~ao, 1010, \\ S\~ao Paulo, SP 05508-090 Brazil 
\\ \vspace{0.5cm}
$^2$S\~ao Carlos Institute of Physics - DFCM \\
University of S\~ao Paulo \\
Av. Trabalhador S\~ao-Carlense, 400, \\ S\~ao Carlos, SP 13566-590 Brazil
}

\date{\emph{30th July, 2024}}

\begin{document}

\maketitle

\begin{abstract}
Different types of graphs and complex networks have been characterized, analyzed, and modeled based on measurements of their respective topology. However, the available networks may constitute approximations of the original structure as a consequence of sampling incompleteness, noise, and/or error in the representation of that structure. Therefore, it becomes of particular interest to quantify how successive modifications may impact a set of adopted topological measurements, and how respectively undergone changes can be interrelated, which has been addressed in this paper by considering similarity networks and hierarchical clustering approaches. These studies are developed respectively to several topological measurements (accessibility, degree, hierarchical degree, clustering coefficient, betweenness centrality, assortativity, and average shortest path) calculated from complex networks of three main types (Erd\H{o}s–Rényi, Barabási–Albert, and geographical) with varying sizes or subjected to progressive edge removal or rewiring. The coincidence similarity index, which can implement particularly strict comparisons, is adopted for two main purposes: to quantify and visualize how the considered topological measurements respond to the considered network alterations and to represent hierarchically the relationships between the observed changes undergone by the considered topological measurements. Several results are reported and discussed, including the identification of three types of topological changes taking place as a consequence of the modifications. In addition, the changes observed for the Erd\H{o}s–Rényi and Barabási–Albert networks resulted mutually more similarly affected by topological changes than for the geometrical networks. The latter type of network has been identified to have more heterogeneous topological features than the other two types of networks.
\end{abstract}

\section{Introduction}\label{sec:introduction}

Complex networks, the main subject of \emph{network science} (e.g.~\cite{barabasi2013network, newman2018networks}), are characterized by diverse types of heterogeneous interconnections between their respective nodes. For instance, scale-free networks tend to have hubs, while geographical structures are characterized by relatively large average shortest path length. Interestingly, networks with distinct topological organization may potentially impact on dynamics taking place on these structures (e.g.~\cite{newman2003structure,boccaletti2006complex,costa2007characterization}). A substantial part of the interest in network science thus relates to the characterization of these topological properties respectively to different types of networks and associated parameter configurations.

Among the several investigations that have been developed aimed at characterizing the topology of networks, some studies (e.g.~\cite{luscombe2004genomic,costa2004reinforcing,boccaletti2006complex, latapy2008complex,bettencourt2009scientific,aerts2016brain,casali2019topological}) addressed the issue of how a given set of topological properties of a given complex network can change as a consequence of topological alterations including changes of network size, edge or node removal, addition, or rewiring. This type of research is of particular interest because a better basic knowledge about this problem can provide insights on how topological or dynamic overall properties can change in the presence of network alterations. Networks can have their structure changed not only as a consequence of failure/attack but also because of nodes and edges of complex real-world structures not being properly identified and taken into account into their respective modeling in terms of complex networks. For instance, it is a particular challenge to fully identify and account for every interactions taking place in specific problems in areas including biology (e.g.~\cite{de2005complex,may2006,buchanan2010}), neuroscience (e.g.~\cite{reijneveld2007,meunier2010,rodrigues2016}), urban sciences (e.g.~\cite{jiang2004,crucitti2006,bettencourt2021}), transportation (e,g,~\cite{zanin2013,lin2013,soh2010}), and ecology (e.g.~\cite{bascompte2007,proulx2005}).

Another aspect of particular interest regards how connectivity of a given type of network (i.e.~its number of edges) may influence its topology, as reflected by respective measurements. This type of situation may occur as a consequence of a larger network not being fully considered, or when networks of a same type can be generated with the same parameter configuration but with varying sizes. Though related to network sampling, the study of how measurements may change with network size is also important for better understanding the effects of \emph{finite size effects} on the respective topology (e.g.~\cite{boguna2004,hong2007,dorogovtsev2008}). 

The present work develops along the above-discussed motivation, i.e.~studying how the topological properties (namely accessibility, degree, hierarchical degree, clustering coefficient, betweenness centrality, assortativity, and average shortest path) of three types of complex networks, namely Erd\H{o}s–Rényi (ER)~\cite{erdos1959random}, Barabási–Albert (BA)~\cite{barabasi1999BA}, geographical (GEO)~\cite{benatti2023simple}, undergone alterations as the network size is changed or the edges of a network under study are progressively remover or rewired. 

In particular, we aim at developing a study of the changes undergone by measurements as a consequence of respective network changes in a comprehensive manner which can complement previous related studies in the following ways: (i) include other measurement such as the node accessibility; (ii) consider the way (shape) in which the measurements change in relative terms (instead of using the respective magnitudes of change); (iii) employ coincidence similarity networks (e.g.~\cite{da2022coincidence}) to interrelate the way in which the measurements respond to network changes of varying types; and (iv) adopt a similarity-based agglomerative clustering~\cite{benatti2024agglomerative}) so as to quantify an hierarchically interrelate the measurement changes.  Several results are described and discussed, corroborating the potential of employing the multiset-based approaches to comprehensive and systematic analysis of complex network topology while emphasizing the study of interrelationships between the measurements changes.

The present work starts by presenting the basic concepts and methods used in the reported approaches and follows by presenting the obtained results and respective discussions.

\section{Related Works and Motivation}

The effect of network changes on the respective topology, as quantified by respective measurements, has been studied from many perspectives jointly considering some type of network alteration and topological characterization of the respective networks. These include studies on network resilience to attacks (e.g.~\cite{shekhtman2015resilience,gao2016,laishram2018measuring}), sampling networks (e.g.~\cite{guillaume2005relevance,dall2005statistical,mahadevan2006internet,serrano2007decoding,frantz2009}), and growing networks (e.g.~\cite{newman2001,dorogovtsev2001,vazquez2003}), as well as works addressing in a more generic manner how progressive network changes can influence its topology, some of which are briefly revised in the following.

A study of how the resilience (a property directly related to the topology of networks) can be influences by the addition of edges to the given networks has been presented in~\cite{costa2004reinforcing}.  Among the considered network changes, edges were added according to an augmentation scheme involving the completing of triangles of edges. While random networks were identified to tend to be more resilient than scale-free counterparts, both these types of responded in similar manners to the considered reinforcement schemes.

In~\cite{boas2010sensitivity}, the sensitivity of 10 topological measurements to topological alterations involving addition, removal, and rewiring of edges are described respectively to four types of networks. Among the several findings reported in that work, the rewiring of edges was found to have a more prominent effect on the network topology than the other two types of changes. In addition, geographical networks were found to be particularly susceptible to the considered types of edge perturbations. The most resilient measurements were identified as corresponding to the node degree, degree of the nearest nodes, hierarchical degree for $h=2$, and clustering coefficient.

The present work aims at developing a more comprehensive study of the effects of network changes on respective topological measurements. This complementation involves: (i) considering more measurements, especially the node accessibility in both its multiscale and generalized versions; (ii) consideration of the way in which the measurements change, which is achieved by using normalized features instead of the original magnitudes of changes (such as adopted in~\cite{boas2010sensitivity}); (iii) use similarity networks to interrelate the way in which the measurements take place; and (iv) employment of similarity-based agglomerative hierarchical clustering in order to quantify the interrelationship between the measurements and groups of measurements in terms of dendrograms.

In particular, the coincidence similarity index is adopted for obtaining the similarity networks, therefore allowing enhanced ability to perform particularly strict comparisons allowing differentiation between similar patterns and, consequently, more detailed similarity networks. Similarity networks have several intrinsic interesting properties because they represent directly how the measurements are related regarding their response to network changes. In addition to observing pairwise interactions, groups can also be formed, indicating that subsets of the considered measurements change in a particularly similar manner (the pairwise similarity can be visualized as the widths of the links in the network). This feature was of particular interest in the present work, allowing the identification of three types of measurements clustered into respective modules. By providing a hierarchical representation of how the measurement changes are interrelated, the adopted hierarchical clustering approach allows a complementation of the information provided by the respective similarity networks, including the possibility to take into account the lengths of the branches of the obtained dendrograms as an indication of the relevance of the respective modules. In addition, the heights of the mergings of pairs of subgroups in dendrograms provide an objective quantification of the respective interrelationship that can effectively complement the results obtained from the similarity networks. Another possibility of particular interest that has been developed in this work is to use the information provided by the obtained similarity networks and respective dendrograms to obtain comprehensive maps indicating how the measurement membership respectively to the identified groups change for each type of network and network variation experiment.

\section{Concepts and Methods}\label{sec:methods}

\begin{figure}
  \centering
     \includegraphics[width=0.9 \textwidth]{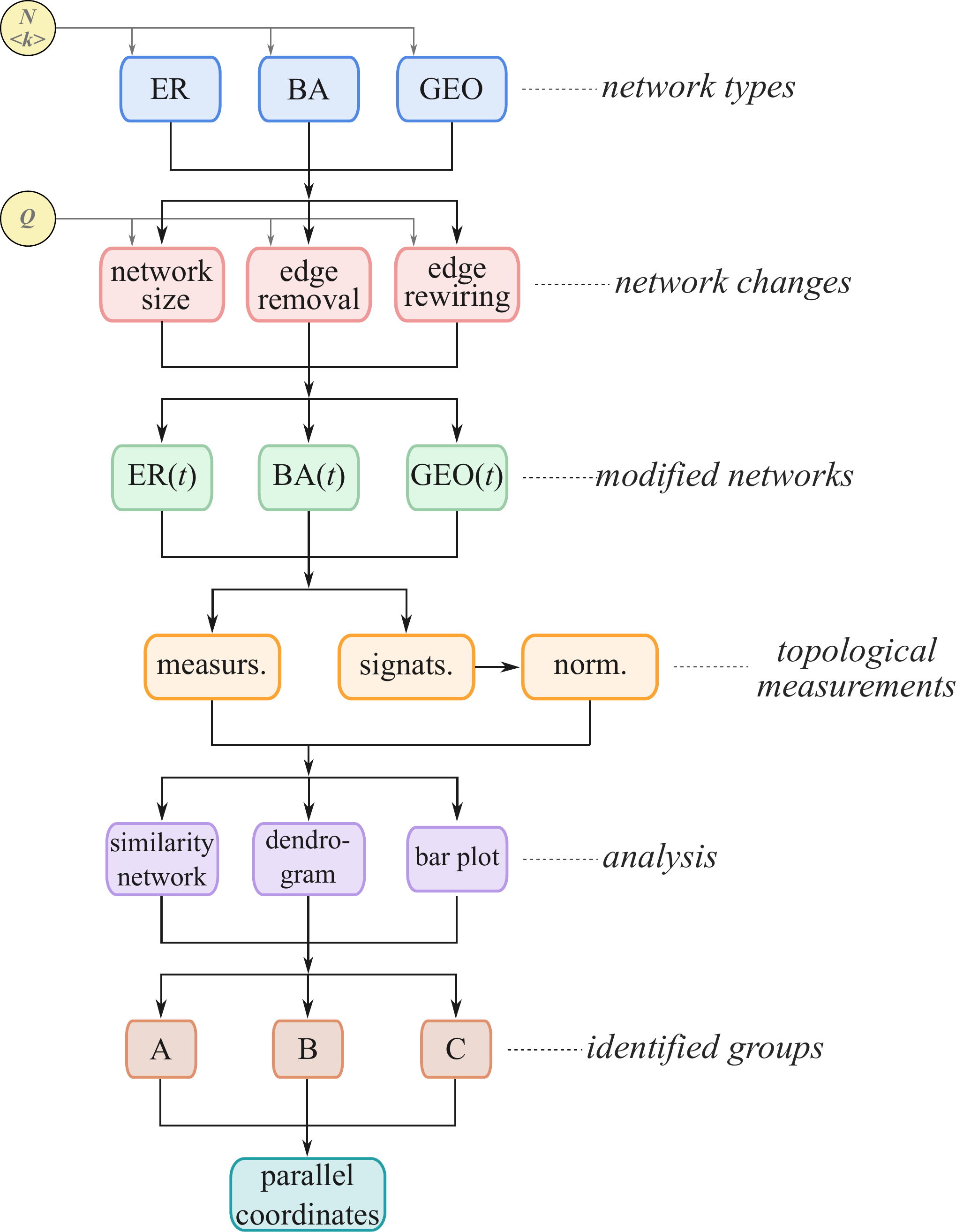} \\
 \caption{Flow diagram illustrating the main stages of the adopted methodology. First, networks of types ER, BA, and GEO are generated, for respective configurations involving the number of nodes $N$ and the average degree $\langle k \rangle$ of the networks. These networks are then subjected to progressive perturbations related to variation of network size, edge removal, and edge rewiring, yielding new versions of the original networks. Measurements of the topology of these networks and used to obtain respective similarity networks, dendrograms, and bar plots of several properties of the measurement changes. Three main types of modifications undergone by the changing measurements as a consequence of the topological changes have been identified, allowing the construction of parallel coordinates plots summarizing the several aspects of how the considered measurements changed in consequence of the considered three types of perturbations.}\label{fig:flow_diagram}
\end{figure}

Several concepts and methods considered in the present work are briefly described in this section.  Figure~\ref{fig:flow_diagram} depicts a flow diagram of the adopted methodology.

\subsection{Complex Networks Generation}\label{subsec:network_generation}

Though the three types of networks adopted in the present work have been described before, they are reviewed in some detail here respectively to the experiments involving networks of the same type and parameter configuration, but with varying sizes. Particular attention is given to how the average node degree is expected to vary with the network size. The ER and BA networks have been obtained considering~\cite{igraph,the2006csardi,barabasi1999emergence}.

In the present work, ER networks have been generated by randomly distributing a given number of edges (implied by the respectively specified average node degree $a_k$) between all possible pairs of distinct nodes among the total number of nodes $N$, which corresponds to the network size. As a consequence, the average node degree is kept fully constant irrespectively of the network size.

Scale-free complex networks of the BA type have been obtained by considering the total number of nodes $N$ and the number of edges $m$ to be incorporated with each new node. The average node degree therefore can be estimated as $a_k \approx 2m/N$. The networks are generated by using a prefix-sum tree, yielding no multiple edges. In the case of not so large networks, the adopted approach to generate BA networks of increasing sizes tend to lead to the average degree also to increase.

The geographical networks (GEO) studied in the present work have been obtained as follows. First, a set of $n \times n$ points is distributed as a non-periodical orthogonal lattice within a two-dimensional square space.  Each of these points is understood as a network node, which yields a total of nodes (network size) equal to $N=n^2$. Small perturbations $\varepsilon$ are then added independently to both coordinates of each of the $n^2$ points, and the respectively obtained Delaunay tessellation (e.g.~\cite{riedinger1988delaunay})] is taken as the resulting geographical network. The structures generated in this manner are said to be geographical not only because their nodes have coordinates, but also because nodes that are closer one another have higher changes of interconnecting. The value of $\varepsilon$ controls the regularity of the obtained networks, with a regular networks (except at the borders) being obtained for $\varepsilon=0$. In the present work, the GEO networks have been obtained with $\varepsilon=0.001$. It is interesting to observe that average node degree of the above type of GEO networks effectively obtained will tend to increase with the network size respectively to the initially specified value $k_a$. This is because the nodes at the border of the network tend to have smaller degrees, implying on only networks of infinite size converging to the value $k_a$=6. The adoption of this specific type of geographical network has been motivated by its intrinsic conceptual simplicity as well as because the links between nodes are defined by their adjacency, therefore strongly reinforcing its geographical nature.

Though not accounting for all possible complex networks topologies, the three considered models descried above still provide a comprehensive and complementary set of interconnectivity patters (including uniformly random, scale free with presence of hubs, small and large average shortest path lengths, as well as varying values of local clustering) intrinsic to several theoretical and real-world networks.

\subsection{Topological Measurements}

The topology of complex networks can be characterized by using a set of respective distinct and complementary measurements. Topological measurements can be subdivided into two large groups: \emph{local} and \emph{global}. The former comprehends the measurements taking into account only the edges connected to a specific node, or within a small neighborhood around that node. Examples of local measurements include the node degree and clustering coefficient. On the other hand, global measurements take into account the whole network or a large neighborhood around a node of interest. Statistical indices (e.g.~average and standard deviation) of local measurements taken on the whole network, as well as the average shortest path length, constitute examples of global topological measurements of complex networks. A third category can be posited, corresponding to an intermediate, mesoscopic scale corresponding to neighborhoods of intermediate sizes. 

Measurements of the interconnectivity of complex networks can actually be extended to \emph{topological multiscale} approaches, with neighborhoods with progressive sizes being considered around a given node. Examples of these type of measurements include (but are not limited to) the hierarchical degree, hierarchical clustering coefficient, and accessibility, all of which are estimated along successive neighborhoods centered at a given node.

\emph{Hierarchical quantifications} of the topological property of networks (e.g.~\cite{newman2003ego,trusina2004hierarchy,da2006hierarchical,costa2007characterization}) respectively to given nodes have been used to characterize how the network topology changes within successive neighborhoods around a give reference node. These neighborhoods are here specified by the parameter $h$, corresponding to the topological distance between the reference node and each successive neighborhood. More specifically, the set of network nodes that are at topological distance $h$ from the reference is henceforth understood as the $h-$neighborhood of that node. Here, the immediate neighborhood of a node is indicated by $h=1$. The set of nodes in the neighborhoods of a given node up to hierarchical level $h$ are called the $h-$ ball of that node. Hierarchical measurements can then be estimated by adapting more traditional topological features such as the node degree and clustering coefficient in terms of the reference node and respective neighborhoods or balls. In the present work, we adopt the hierarchical degree for hierarchical levels $h$ varying from $=2$ to $5$.

The concept of \emph{node accessibility}, which applies to a given reference node, has been described (e.g.~\cite{travenccolo2008accessibility,benatti2022complex}) as a means to quantify the number of nodes that are effectively accessed while performing some dynamics on the network. In the present work, this dynamic is assumed to correspond to uniform random walks at the equilibrium regime. The effective number of nodes that can be accessed after $h$ steps by an agent departing from the reference node can then be taken as corresponding to the exponential of the entropy of the transition probabilities extending from the reference node toward each of the nodes in its $h-$neighborhood. It can be show that the accessibility for $h=1$ in an unweighted network corresponds to the degree of the reference node. For this reason, the accessibility has been understood as a possible generalization of the node degree (e.g.~\cite{benatti2022complex}). A directly related measurement, namely the \emph{generalized accessibility}~\cite{de2014role,benatti2022complex} has also been considered in the present work. This measurement corresponds to a generalization of the accessibility in order to take into account all possible hierarchical levels (up to infinity), considering all paths vising the reference node.

Observe that, though local measurements are considered, their global average taking into account the whole network is taken into account in the reported experiments.

The measurements of complex networks topology considered in the present work are listed in Table~\ref{tab:meas}, which also indicates their adopted abbreviation and type (local, global, mesoscopic).

\begin{table}
\centering
\begin{tabular}{ |l|c|c| } 
 \hline
 \textbf{\emph{measurement}} & \textbf{\emph{abbreviation}} &  \textbf{\emph{type}}   \\ \hline  \hline
 Average node degree & \emph{Degree} & local  \\   \hline
 Clustering Coefficient &  \emph{Clust.~Coeff.}  & local \\   \hline
 Betweenness centrality &  \emph{Betw.~Centr.}  & global \\   \hline
 Assortativity &  \emph{Assortativity}  & global \\   \hline
 Average Shortest Path Length &  \emph{Avg.~Short.~Paths}  & global \\   \hline
 Hierarchical degree for $h=2$ &  \emph{Hier.~Deg\_h2}  & mesoscopic \\   \hline
 Hierarchical degree for $h=3$ &  \emph{Hier.~Deg\_h3}  & mesoscopic \\   \hline
 Hierarchical degree for $h=4$ &  \emph{Hier.~Deg\_h4}  & mesoscopic \\   \hline
 Hierarchical degree for $h=5$ &  \emph{Hier.~Deg\_h5}  & mesoscopic \\   \hline
 Accessibility for $h=2$ &  \emph{Access\_h2}  & mesoscopic \\   \hline
 Accessibility for $h=3$ &  \emph{Access\_h3}  & mesoscopic \\   \hline
 Accessibility for $h=4$ &  \emph{Access\_h4}  & mesoscopic \\   \hline
 Accessibility for $h=5$ &  \emph{Access\_h5}  & mesoscopic \\   \hline
 Generalized Accessibility &  \emph{Gen.~Access.}  & global \\   \hline
 
\end{tabular}
 \caption{The topological measurements of complex networks considered in the present work as well as their respective abbreviations and types.}\label{tab:meas}
\end{table}

\subsection{Normalization}

Henceforth, we consider that each of $N$ data elements is characterized in terms of $M$ respective features indexed by $j=1, 2, \ldots, M$. Therefore, each data element $i$ can be represented in terms of a respective \emph{feature vector} $\vec{x}_i = [x_{i,j}]$.

Once a representative set of topological measurements has been obtained from a given complex network, it is typically interesting to organize and further study them in order to better understand the structure of the networks of interest. Because distinct measurements can have markedly distinct ranges of values (e.g.~the node degree can vary from 1 to dozens or hundreds, while the clustering coefficient is comprised in the interval $[0,1]$), it is often interesting to normalize the available set of values in some related way.  

The choice of a specific normalization approach can strongly influence subsequent analysis and classification of the data.  Special attention needs to be given to the type of available data as well as to how they will be analyzed.  In principle, proportional comparisons are intrinsically related to proportional data~\cite{propnorm}.

In the present work, all considered measurements have nonnegative values, but different ranges of respective magnitudes.  Each of the available features $j$ corresponds to the set of values $x_{i,j}$, $i=1, 2, \ldots, N$, of one of the topological measurements considered taken along the progressive changes of a complex network $i=1, 2, \ldots, N$.  The variation of these measurements is henceforth quantified in terms of signatures $s_{i,j} = x_{i,j} - \min(x_{i,j})$, so as to have the same baseline.  In order to achieve signature ranges with comparable magnitudes, the values of each of the signatures were normalized by dividing by the respective standard deviation, i.e.:
\begin{equation}\label{eq:norm}
    c_{i,j} = \frac{{s}_{i,j}}  {\sigma_j} = \frac{{x}_{i,j} - \min \left(x_{i,j}\right)}  {\sigma_j}
\end{equation}  
where $\mu_j$ and $\sigma_j$ are respectively the average and standard deviation of each of the features $j$, considering the whole dataset.  

The resulting normalized features $\tilde{x}_{i,j}$ have non-negative values and unit standard deviation.    In addition, observe that the resulting normalized values are dimensionless.

\subsection{Multiset Similarity Indices and Networks}

Though comparisons between measurements and vectors have been often performed in terms of \emph{distances}, the present work focus on complementing these approaches by considering instead the \emph{similarity} between two measurements or vectors (multisets). More specifically, we resource to the \emph{coincidence similarity index} (e.g.~\cite{da2021further, costa2023mneurons, benatti2024agglomerative}), which can be understood as the Jaccard similarity index modified in order to take into account the relative interiority between the compared data.

The Jaccard similarity index (e.g.~\cite{Jac:wiki,wu2018,da2021further,rozinek2021}) is on itself a particularly simple and conceptually appealing approach to quantifying the similarity between two non-empty sets in terms of the set (and multiset) operations of intersection and union.  However, as discussed in~\cite{da2021further}, the Jaccard index does not take into account the relative interiority between the compared sets.  This limitation can be addressed~\cite{da2021further} by multiplying the Jaccard index by the interiority (or overlap,~\cite{vijaymeena2016a}) index, yielding a measurement of similarity which has been called \emph{coincidence similarity index}. 

The \emph{coincidence similarity index} considering two $M-$dimensional vectors $\vec{x} = [x_i]$ and $\vec{y} = [y_i]$, both having all elements with positive real values, can be expressed as:
\begin{align}
  \mathcal{C}(\vec{x},\vec{y}) = \mathcal{J}_D(\vec{x},\vec{y})  \ \mathcal{I}_E(\vec{x},\vec{y})  \label{eq:coinc}
\end{align}
where:
\begin{align}
    \mathcal{J}_D(\vec{x},\vec{y}) = \left[  \frac{\sum_i \min(x[i], y[i]) + \delta}{\sum_i \max(x[i], y[i]) + \delta} \right]^D ,
\end{align}
is the multiset Jaccard index,  $\delta > 0$ is a small-valued regularization constant (possibly used to smoothing the comparisons near the singularity $0/0$), and $D$ is a coefficient controlling how strict the implemented similarity comparison is.

The multiset interiority index can be expressed as:
\begin{align}
    \mathcal{I}_E(\vec{x},\vec{y}) =  \left[ \frac{\sum_i \min(x[i], y[i]) + \delta}{\min \{ \sum_ix[i], \sum_i y[i] \} + \delta}  \right]^E.  \label{eq:Jacc}
\end{align}
where $E$ is a parameter controlling how strict the comparison is.

We necessarily have that $0 \leq \mathcal{C}(\vec{x},\vec{y}) \leq 1$, and $\mathcal{C}(\vec{x},\vec{y}) = \mathcal{C}(\vec{y},\vec{x})$.

\begin{figure}
  \centering
     \includegraphics[width=0.9 \textwidth]{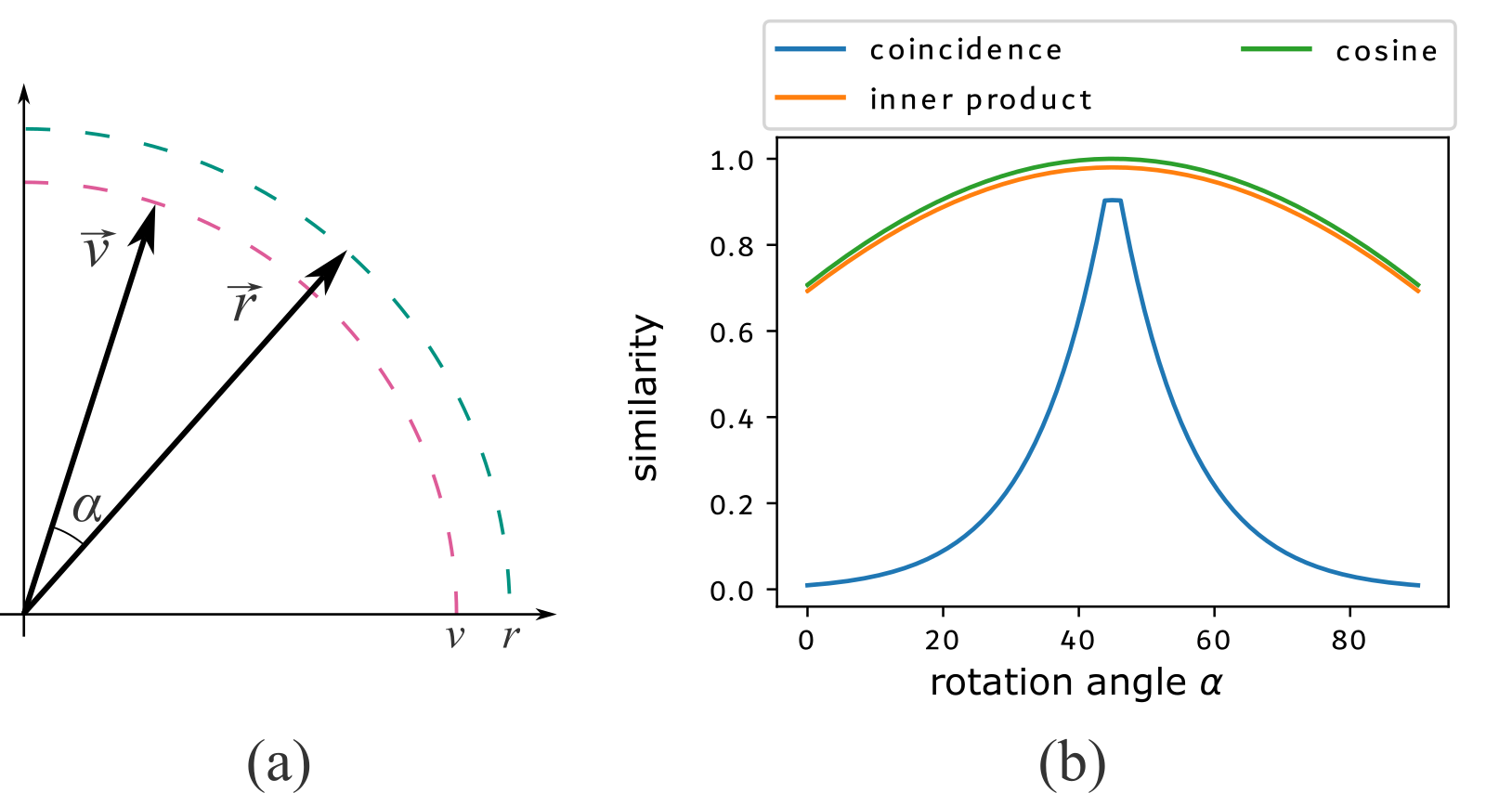}
 \caption{Illustration of the similarity comparison implemented by the coincidence similarity index relatively to three alternative approaches, namely inner product and cosine similarity index. A vector $\vec{r}$ with coordinates $[\sqrt{2}/2,\sqrt{2}/2]$ is compared with vectors $\vec{v}$ with magnitude $0.95$ and varying rotation angles $\alpha \in [0,\pi/2]$, as depicted in panel (a). The coincidence similarity index has been implemented with $\delta = 0$, $D=5$ and $E=1$. The obtained similarity values respectively to varying values of $\alpha$, presented in (b), have substantially smaller magnitudes, leading to more strict comparisons. }\label{fig:similarity}
\end{figure}

The coincidence similarity index has several interesting intrinsic characteristics (e.g.~\cite{da2021further,da2022coincidence,costa2023mneurons}), including being dimensionless, bounded within the interval $[0,1]$, and potential for implementing strict comparisons.  Figure~\ref{fig:similarity} illustrates the comparison of two vectors by using the inner product, cosine similarity, and the coincidence similarity index.

An interesting feature of the coincidence similarity index is its relatively robustness to outliers and limited levels of noise in the data. It should also be kept in mind that the coincidence similarity index (as well as the Jaccard and interiority indices) implements a \emph{proportional comparison}~\cite{benatti2024agglomerative}, in the sense that the comparisons are performed relatively to the magnitude of the compared vectors.  This implies that comparisons between vectors with smaller magnitudes will tend to be more strict than when comparing larger magnitude vectors.  

Given a dataset of $N$ elements, each characterized in terms of $M$ features, it is possible to obtain the coincidence similarity between each possible pair of these elements. These values can be effectively represented and visualized in terms of respective \emph{coincidence similarity networks}~\cite{da2022coincidence}, in which each node corresponds to a data elements while each edge is associated to a weight corresponding to the respective similarity value. The more strict similarity comparison performed by the coincidence similarity index, which becomes even more strict for larger values of $D$ and $E$, contributes not only for better separating similar data elements, but also to obtaining similarity networks with enhanced levels of detailed interconnectivity.

Similarity networks are of particular interest because they convey, as contrasted with distance approaches, a direct representation of the \emph{similarity relationships} between the considered data elements which are, in the case of the present work correspond to the adopted topological measurements of complex networks. Therefore, similarity networks can be of interest for complementing approaches based on distances between data elements. It should be kept in mind that the specific type of data and respective normalization can influence significantly the respectively obtained similarity values.

\subsection{Hierarchical Agglomerative Clustering}\label{subsec:hierarchical}

In addition to allowing respective networks to be inferred, the quantification of the similarity between data elements, similarity indices can also be employed as a means to obtain \emph{hierarchical clustering} approaches as described in~\cite{benatti2024agglomerative} respectively to an agglomerative implementation. Basically, the subgroups of data elements are successively merged according to the maximum similarity between the currently existing groups, therefore defining respective hierarchies along the relationship between the data elements.  These relationships can then be effectively represented and visualized as respective \emph{dendrograms}.

\subsection{Networks Modifications}

Three types of alterations are considered, namely: changes in network size, removal of edges, and rewiring of edges. Each of these experiments is reported in respective subsections. All coincidence similarity networks have been obtained with $\delta=0$, $D=5$, and $E=1$. 

In the case of the network size and edge removal experiments, the number of edges is adopted as the free variable along which the changes are mapped and plot.  However, the number of rewirings had to be adopted instead in the case of the edge rewiring experiment, because the number of edges remains constant in that case.

\section{Results and Discussion}\label{sec:results}

This section describes the performed experiments and respective discussion. These experiments regard the application of coincidence similarity networks and similarity-based hierarchical agglomerative clustering analysis as resources to characterize and study how a set of network measurements change respectively to three types of progressive modifications of ER, BA, and GEO networks. All considered networks had average node degree equal to $5.7$.

The adoption of the number of edges as free variable has the advantage of allowing the same horizontal axis in the plots to be obtained for the network size and edge removal experiments. Recall that the number of undirected edges $E$, number of nodes $N$ and average degree ($\left< k \right>$) are related as follows:
\begin{align}
  \left< k \right> = 2 \, \frac{E}{N}
\end{align}

\subsection{Network Size}\label{subsec:size}

In this subsection, the issue of how the topological measurements of complex networks change respectively to successive alterations of the network size (i.e.~total number of nodes).  This type of change is closely related to networks growing along time through the incorporation of new nodes and links, while maintaining the same type of structure and average degree. Border effects implied by the finite size of the considered networks can potentially impact on the respective topological measurements. 

Though the average node degree has been kept as constant as possible in the case of the ER networks, variations of this parameter are expected in the case of the BA and GEO models. The reported experiments considered the following set of network sizes $S  = \left\{ 16,25,36,49,64,81,100 \right\}$. These network sizes have been employed as they are intrinsically implied by the generation of the GEO networks (please refer to Section~\ref{subsec:network_generation}).  

A total of $Q=1,000$ realizations have been considered in order to obtain statistics, namely average and standard deviation, of each of the adopted measurements in terms of the number of edges, respectively to the network models ER, BA, and GEO. Each of these realizations refers to a distinct network sampled from the ensemble of the respective model.

\begin{figure}
  \centering
     \includegraphics[width=0.99 \textwidth]{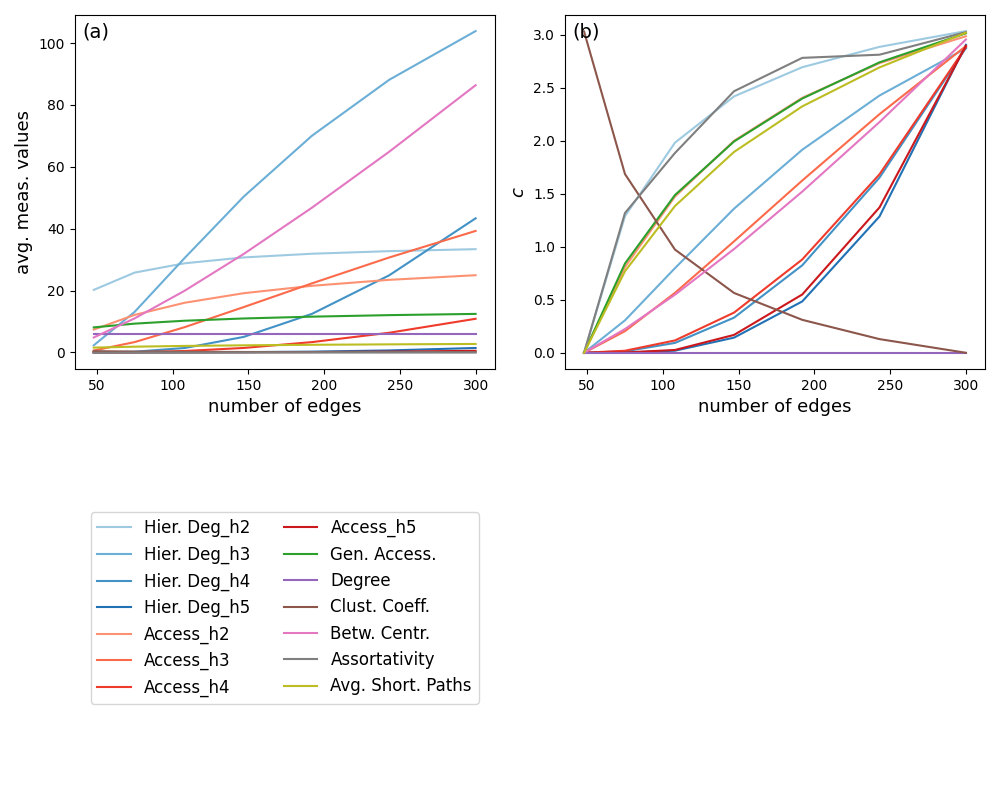}
 \caption{Curves obtained for the measurement changes $\Delta$ in terms of the number of edges involves obtaining the following (over $Q=1,000$ realizations): (a) average of each of the original values obtained for realizations on the ER network model;  and (b) the normalized signatures of the measurements changes, obtained by using Eq.~\ref{eq:norm}, which are considered for the subsequent analysis.}\label{fig:curves_nodes_ER}
\end{figure}

Figure~\ref{fig:curves_nodes_ER} depicts three curves obtained for ER networks. The average of each of the measurements is shown in (a), while the respectively (separately) standardized curves are presented in (b). Each of these curves has null means and unit standard deviation. The measurement changes $c$, shown in (c), have been obtained by subtracting each of the individual curves in (b) by the respective initial value (at network size 16). Because of the adopted normalization, all subsequent comparisons taking into account these curves will reflect the \emph{shape} and \emph{tendency of variation} in these curves, rather than the magnitudes of the original measurements.

Two main groups of curves can be identified: a larger one involving curves with values increasing monotonically with the network size, as well as a group involving a single curve with decreasing values. 

\begin{figure}
  \centering
     \includegraphics[width=0.99 \textwidth]{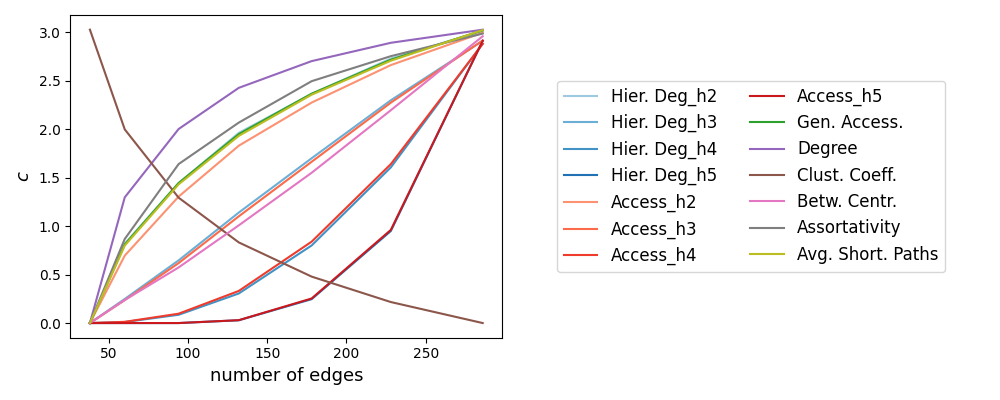}
 \caption{The curve of the measurements changes $c$ in terms of the number of edges respective to the BA network model, containing two main groups as in the ER case.}\label{fig:curves_nodes_BA}
\end{figure}

\begin{figure}
  \centering
     \includegraphics[width=0.99 \textwidth]{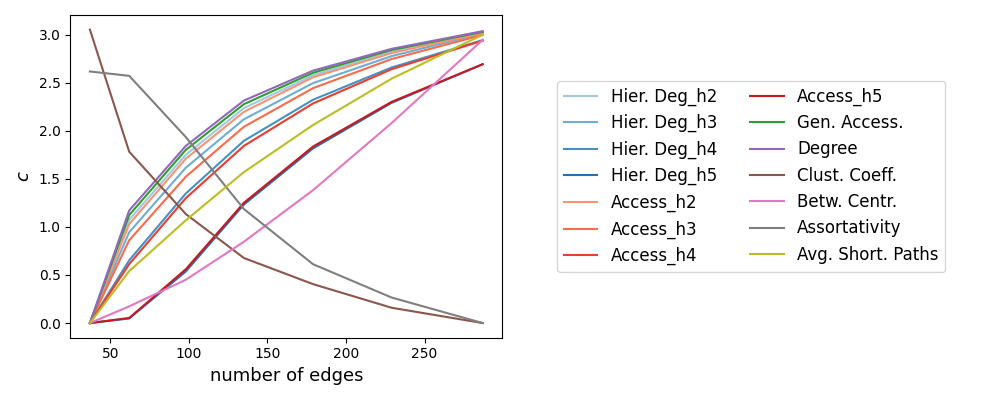}
 \caption{The curve of the measurements changes $c$ in terms of the number of edges respective to the GEO network model, containing two main groups characterized by curves respectively increasing and decreasing with the network size.}\label{fig:curves_nodes_VR}
\end{figure}

Figures~\ref{fig:curves_nodes_BA} and~\ref{fig:curves_nodes_VR} present the curves of $\Delta$ estimated for the experiments respective to the BA and GEO models.
 
In the case of the ER networks, regarding how the measurements change as the network size is increases, we have that the clustering coefficient decreases, the degree remains constant (as could be expected), while the other measurements increase, three of them in nearly linear manner.

It can be observed from the above results that, though the average degree could have been expected to remain constant as the size of the networks increases, this has been more accurately verified only in the case of the ER model.  In the cases of the BA and GEO networks, the average node degree presented an increase with the number of edges. That is because these models impose limitations on the values of average node degree that can be obtained in practice. Therefore, the results henceforth reported and discussed regarding the BA and GEO networks in the case of the network size experiment are not totally specific to the average node degree, which tends to increase with the network size in those cases. 

In addition to paying special attention to the shape of the curves obtained for $\Delta$, as a means of identifying their basic properties, it is also necessary to consider the dispersion of those relationships as observed among the several performed realizations. In this work, we quantified the overall dispersion of each set of curves obtained for each same measurement in terms of an index $\psi$ corresponding to the average of the standard deviations of standardized measurement differences estimated along the considered number of edges values. More specifically, if $R$ is the total of instances considered for the number of edges for each measurement, we could write:
\begin{align}
  \psi = \frac{1}{R} \, \sum_{k=1}^{R} \sigma_k,
\end{align}
where $\sigma_k$ is the standard deviation respective to one of the considered number of edges. 

Thus, a small value of $\psi$ observed for a given measurement respectively to a given network model indicates a well-defined relationship between that measurement and the number of edges $E$. Larger variations of the index $\psi$ observed for a specific curve would thus mean that the respective measurement changes depend on other topological factors.

The estimation of $\psi$ also allows the consideration of eventual specificity to other types of functional relationships, in the sense that a relatively small value of $\psi$ for a given measurement indicates a well-defined relationship between the changes of that measurement and the considered free variable. Although mutual information could have been eventually applied instead of the index $\psi$, it could be influenced by the implied choice of binning schemes.

Because the results and analysis of measurement changes performed in this work focus on the \emph{relative} measurements changes, it may happen that large dispersions obtained for these relative changes actually correspond to originally small, or even immaterial, magnitude variations or statistical fluctuations. Observe that the normalization approach can amplify the dispersion of a set of values originally presenting standard deviations smaller than 1. Thus, it becomes interesting to incorporate also information about the overall \emph{absolute magnitude} of the measurement changes prior to normalization. This is henceforth addressed by taking into account an additional barplot indicating the average of the absolute values of the original, non-standardized, measurement changes. This index is indicated as $M$.

\begin{figure}
  \centering
     \includegraphics[width=0.9 \textwidth]{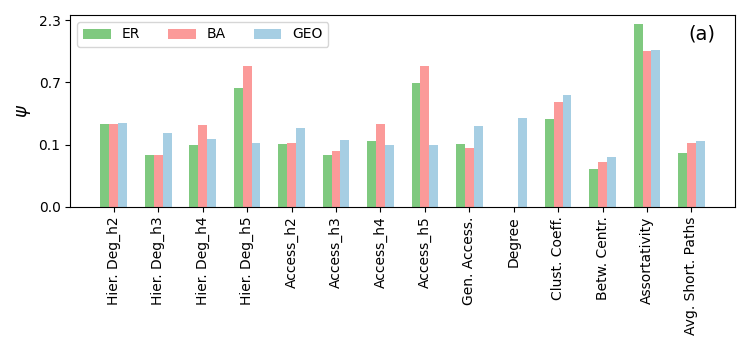}
     \includegraphics[width=0.9 \textwidth]{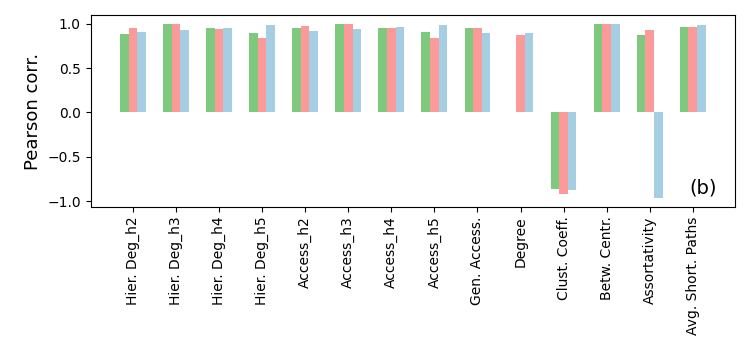}
     \includegraphics[width=0.9 \textwidth]{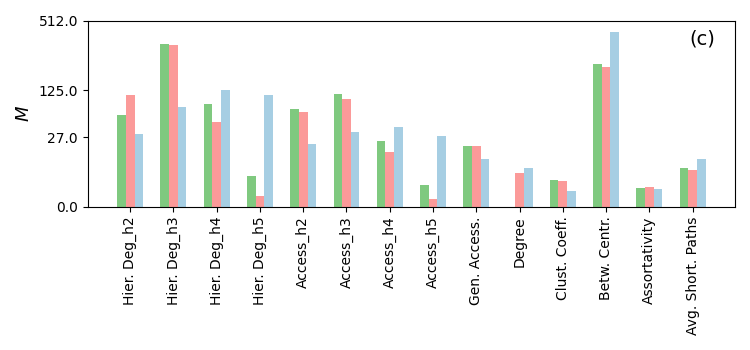}
 \caption{The indices estimating the overall variation $\psi$ (a), the Pearson correlation (b), and the overall magnitude of the non-standardized measurement changes (c) obtained for each of the curves in the network size experiment. The measurements with smaller overall variations can be understood to follow the number of edges in a more specific manner. At the same time, large Pearson correlations indicate a more linear relationship between each measurement and the number of edges. Observe that the vertical axes are presented in non-linear scales in the case of the barplots in (a) and (c) for the sake of enhanced visualization.}\label{fig:barplot_nodes}
\end{figure}

The bar plot in Figure~\ref{fig:barplot_nodes}(a) presents the values of $\psi$ obtained for each measurement and each network model in the case of the network size experiment. It can be noticed that comparable and relatively small overall variations have been observed for the network size experiment.

Actually, the linearity of the curves provide one of the most basic indications about the measurements changes. More specifically, if a curve is close to being linear, this indicates \emph{full correlation} between those two properties. However, full correlation does not necessarily means that there is a causality from the number of edges to each measurement undergoing linear changes, but it does reveal that they are closely related.

Figure~\ref{fig:barplot_nodes}(b) depicts the Pearson correlation of the measurement changes $c$ obtained for each of the network models in the network size experiment. Relatively high absolute values can be observed for several of the measurements, with a few exceptions. When taken jointly with the respective values of $\psi$, we have that specific relationships have been obtained between the measurement changes and the number of edges. This result indicates that, in this type of experiment, most of the relative measurement changes quantified by $\Delta$ are directly related to the number of edges or, more informally speaking, it could be said that the number of edges is the main aspect determining how the several considered measurements change when the size of the networks is progressively modified.

The overall magnitudes $M$ of the non-standardized measurements shown in the barplot in Figure~\ref{fig:barplot_nodes}(c) reveal substantially smaller magnitudes for the measurement changes in the case of the degree, clustering coefficient, and assortativity, while the betweenness centrality was characterized by larger magnitude changes.  The assortativity presents the largest values of the overall variation $\psi$ among the realizations.

\begin{figure}
  \centering
     \includegraphics[width=0.99 \textwidth]{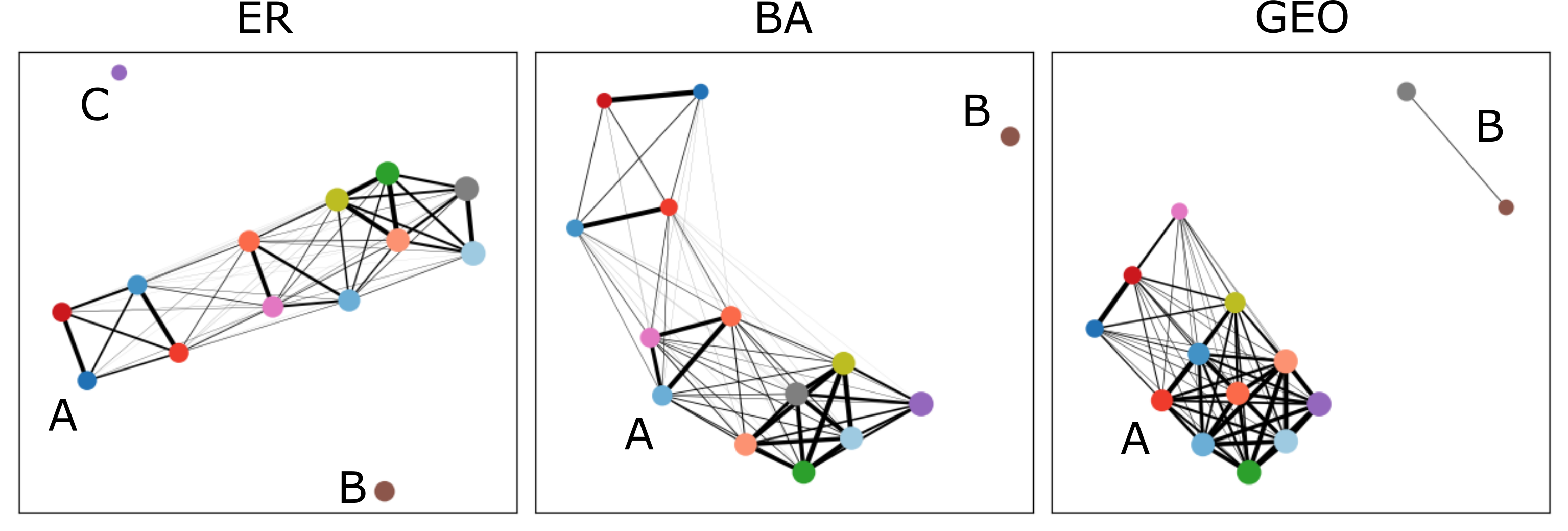}\\
     \hspace{.1cm} (a) \hspace{3.6cm} (b) \hspace{3.6cm} (c)
 \caption{The coincidence similarity networks obtained for the experiments involving varying network sizes respectively to the following network models: (a) ER, involving three modules; (b) BA, involving two modules; and (c) GEO, involving two components. The color scheme follows the same convention presented in Fig.~\ref{fig:curves_nodes_ER}. The width of the edges are proportional to the coincidence similarity index between the respective pair of measurements. The size of the nodes provide an indication of the overall relative change undergone by that node considering the whole set of network sizes.
 }\label{fig:net_nodes}
\end{figure}

The estimated values of the measurements changes $c$ performed as described above have also been used in order to obtain coincidence similarity complex networks for each of the experiments, namely considering ER, BA, and GEO models, which are presented in Figure~\ref{fig:net_nodes}.  More specifically, pairs of the obtained curves $c$ are understood as the vectors $\vec{x}$ and $\vec{y}$ in Equation~\ref{eq:coinc}. The size of the nodes are proportional to the total area of absolute values of the curves obtained for $c$. Therefore, the larger the size of the nodes, the more intense the displacement of the respective measurement respectively to the value it had in the initial network.  In addition, the width of the edges have been drawn proportionally to the respective coincidence similarity index, so that a wider edge between two measurements indicate that they are particularly similar.

In the case of all three considered network models, the networks yielded well-defined modules which resulted markedly separated one another in the respective visualizations. These groups can be verified to correspond to curves $\Delta$ that mostly increase, decrease or have another type of variation (e.g.~remain constant or oscillate substantially). For the sake of effective referencing the modules resulting from the obtained coincidence similarity networks, the following convention is henceforth adopted: measurements mostly undergoing monotonical increase as a consequence of the implemented topological changes are labeled as $A$, while monotonically decreasing measurements are labeled as $B$. Measurements undergoing other types of changes (e.g.~remaining constant or presenting substantial oscillations) are labeled as $C$.

Several additional information can be gathered from the obtained coincidence similarity networks. First, the groups previously identified as $A$, $B$, and $C$ from the curves $c$ in Figures~\ref{fig:curves_nodes_ER}, ~\ref{fig:curves_nodes_BA} and ~\ref{fig:curves_nodes_VR}, resulted in markedly separated network modules, as can be visualized in Figure~\ref{fig:net_nodes}. In addition, the substantially varying widths of the edges obtained in the larger groups indicate that, though all these measurements increase with the network size, they do so in largely distinct manners. Interestingly, the placement of the nodes in the obtained coincidence similarity networks follows a well-defined order progressing from the nodes increasing in a superlinear manner toward those increasing sublinearly (left-hand side), with linear increase tending to result in the middle of the respective network modules.

Of particular interest is the fact that only the clustering coefficient and assortativity measurements presented a tendency to decrease with the size of the network, indicating effects of the network finite sizes. The former measurement decreased for all three considered networks types. Both effects are possibly related to the fact that the number of border nodes tends to decrease relatively to the network size. The clustering coefficient decreases with the network size because it is a local measurement that tends to be larger at the network border nodes. The assortativity decreased only in the case of GEO networks. This effect can be related to the fact that the shortest paths in a larger geographical network tend not only to be larger than those in the ER and BA cases, but also because these shortest paths tend to be distributed among the inner network nodes.

\begin{figure}
  \centering
     \includegraphics[width=0.99 \textwidth]{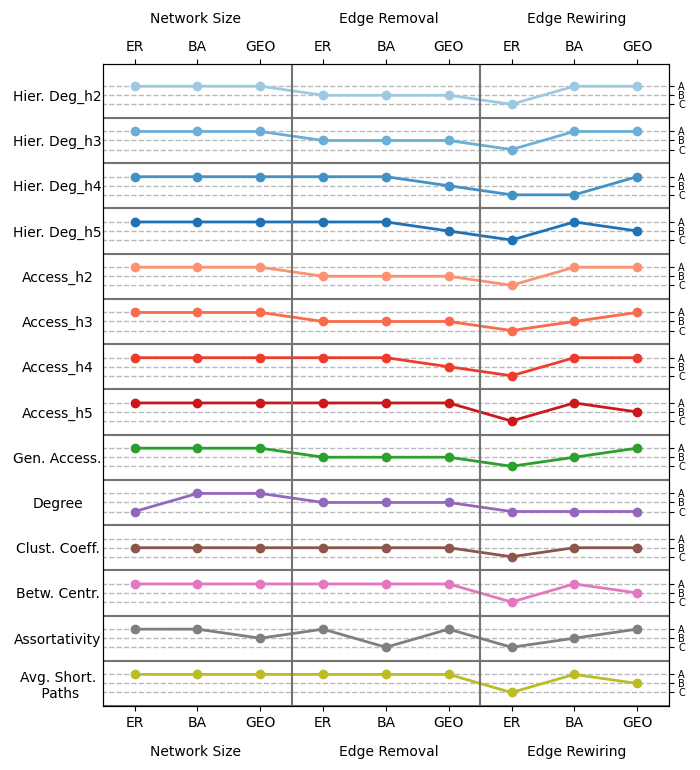}
 \caption{Diagram indicating the changes of category (respectively to the groups types $A$, $B$, and $C$) undergone by each of the considered topological measurements respectively to the three performed experiments and three types of complex networks. The types of changes are indicated in terms of the labels $A$ (mostly monotonical increase), $B$ (mostly monotonical decrease), and $C$ (other types of changes, such as remaining constant or presenting substantial oscillations).}\label{fig:diagram}
\end{figure}

Figure~\ref{fig:diagram} illustrates a diagram showing the membership of each measurement respectively to the three types of progressive modifications (network size, edge removal, and edge rewiring) of the network models ER, BA, and GEO. It can be seen from this diagram that, in the case of the network size experiment, most measurement changes kept into the same overall group, except for the node degree and assortativity.

The heterogeneity of interconnection intensities typically observed from coincidence similarity networks motivates further complementary analysis aimed at organizing these intricate relationships into an effective manner. A particularly interesting possibility is to organize the obtained similarity relationships \emph{hierarchically}, for instance by using hierarchical agglomerative clustering (e.g.~\cite{benatti2024agglomerative}) as well as the respectively derived \emph{dendrograms}, as described in Section~\ref{subsec:hierarchical}.

\begin{figure}
  \centering
     \includegraphics[width=0.99 \textwidth]{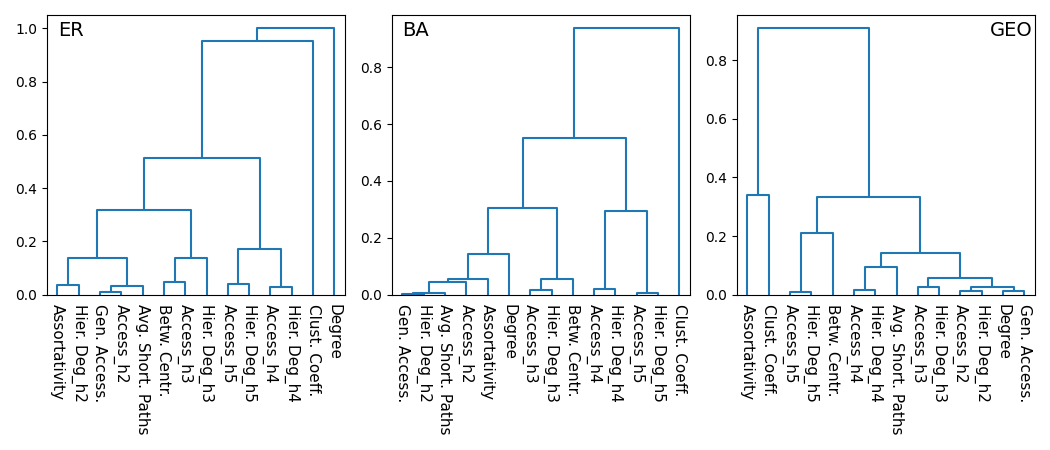}\\
     \hspace{.1cm} (a) \hspace{3.6cm} (b) \hspace{3.6cm} (c)
 \caption{Dendrograms obtained by hierarchical agglomerative clustering of the coincidence similarity networks respectively to the network size experiments (Fig.~\ref{fig:net_nodes}).}\label{fig:den_nodes}
\end{figure}

In this work, we employ a hierarchical agglomerative clustering methodology~\cite{benatti2024agglomerative} based on coincidence similarity relationships between the weights of the coincidence similarity networks obtained for measurements variations. Figure~\ref{fig:den_nodes} illustrates the dendrograms obtained respectively to the experiments, represented as coincidence similarity networks in Figure~\ref{fig:net_nodes}, investigating the effect of network size on ER (a), BA (b), and GEO (c) networks.

The obtained dendrograms present in an effective manner the hierarchical interrelationships between the measurements changes undergone as a consequence of the progressive increase of network size. Of particular interest is the fact that the hierarchical representation of the relationships allows subgroups, or subfamilies, of measurement changes to be identified. 

The dendrograms obtained for the ER and BA cases are mostly similar, with some small differences at the lowest hierarchical levels.  On the other hand, the dendrogram obtained for the GEO networks is markedly distinct from the dendrograms for ER and BA networks. First, we have that the changes respective to the assortativity measurement decrease for the GEO networks and increase for the other two network models. This happens because, with the increase of the GEO network size, the nodes as well as their initial neighbors become progressively more independent of other portions of the network as the latter becomes less intertwined as a consequence of the relatively large values of average shortest path lengths. Another aspect in which the results obtained for the GEO networks differ from the ER and BA models concerns the fact that the subgroups observed for the former type of network are substantially more compact, therefore yielding denser respective modules in Figure~\ref{fig:net_nodes} as well as shorter respective branches in the dendrograms in Figure~\ref{fig:den_nodes}. At the same time, the identified modules are qualitatively closely related among all the three considered networks models. 

All in all, the above reported systematic approach and discussion allowed the identification of several trends in the relationship between the changes of topological networks respectively to the three main adopted models. In addition to several more specific observations, we have that almost none (other than the node degree in the case of the ER model) of the measurements have been kept constant as the network size increases, undergoing respective increase or decrease of values. These results illustrate how challenging the comparison between networks of different types can be.

\subsection{Edge Removal}\label{subsec:removal}

Another type of study involving progressive changes on the topology of complex networks consists of the successive removal of edges, while the number of nodes (network size) remains constant.  The current subsection reports on the experimental results and respective discussion concerning this type of progressive topological alterations while considering the ER, BA, and GEO models. 

One first point to be kept in mind regards the fact that the way in which the number of edges varies in this type of experiment differs in two significant manners from the previous experiment.  First, as edges are removed, the number of edges consequently decreases in a linear manner, implying a respective change in the average degree of the network.  In addition, unlike the previous experiment, the number of nodes (network size) remains constant during the whole edge removal procedure.  By itself, the above observed properties indicate not only that the network size will not impact on how the measurements can change during the experiment, but also that these changes can proceed in a distinct manner as observed in the previous experiment, where the network size \emph{and} number of edges have been made both to progressively increase so as to keep the average node degree as constant as possible.

All networks have their size fixed at 100 nodes, with initial average node degree equal to 5.7. A total of $Q = 50$ realizations were performed respectively to each type of network for the number of removed edges varying from 0 to 100. These realizations correspond to distinct sequences of edge removal performed on a same initial network for each model.

\begin{figure}
  \centering
     \includegraphics[width=0.99 \textwidth]{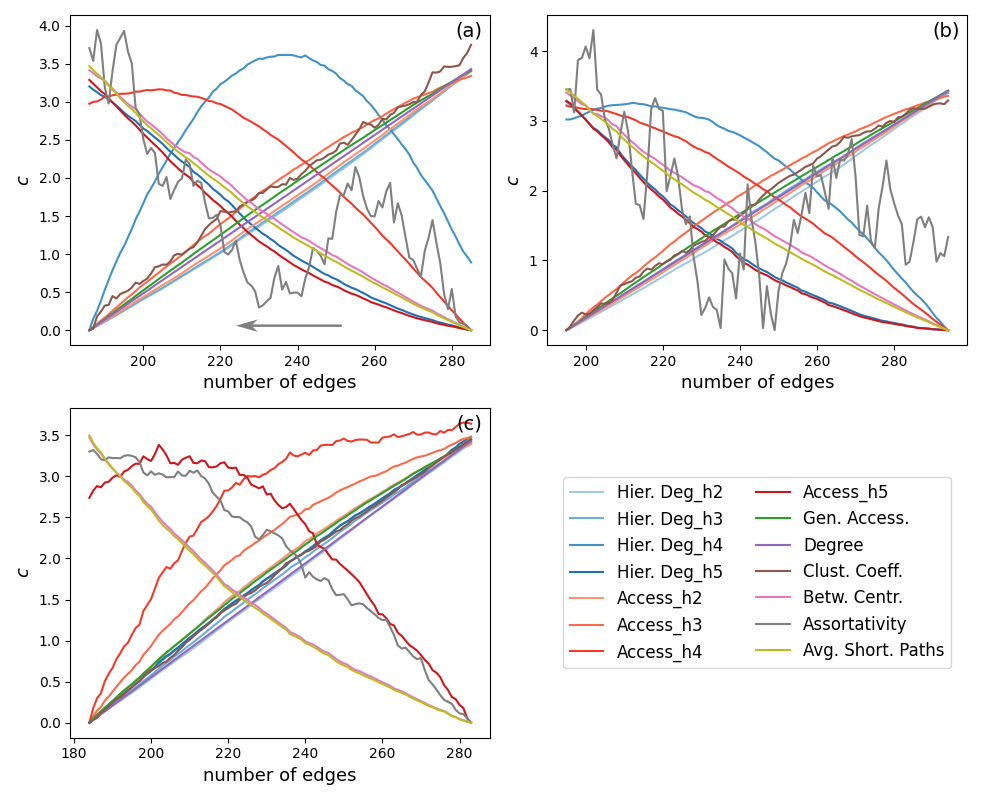}
 \caption{Curves of measurements changes $c$ in terms of the number of progressively removed edges respectively to the ER (a), BA (b), and GEO (c) types of networks. Markedly distinct results have been obtained for each of these cases. The arrow indicate the direction of changes of number of edges implied by the successive removal of edges.}\label{fig:curves_remove}
\end{figure}

Figure~\ref{fig:curves_remove} shows the curves of the measurements changes $\Delta$ in terms of the number of removed edges for the considered ER (a), BA (b), and GEO (c) types of networks. 

In a manner similar to that observed in the network size experiment presented in Section~\ref{subsec:size}, the curves of $c$ were subdivided into two major groups, corresponding to situations in which the respective values increase (group $A$) or decrease (group $B$) in mostly monotonical manner with the number of edges. However, unlike in that experiment, some curves contained a mixture of increasing or decreasing portions, which appeared smoothly or in oscillating and jagged manner. These cases have been understood as corresponding to group $C$. No curve remained nearly constant as the edges were progressively removed. Interestingly, the curves undergoing a monotonical decrease tended to be more linear that the other types of observed curves.

\begin{figure}
  \centering
     \includegraphics[width=0.9 \textwidth]{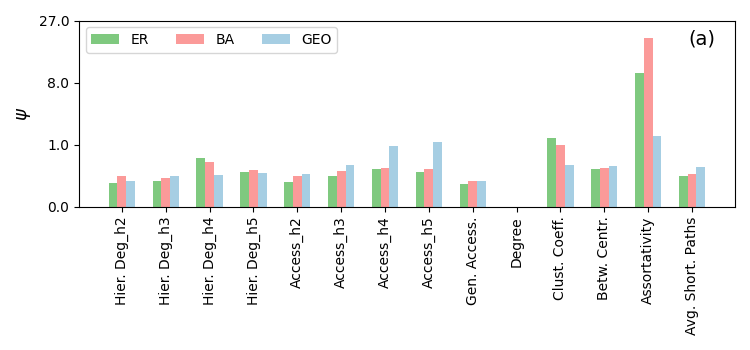}
     \includegraphics[width=0.9 \textwidth]{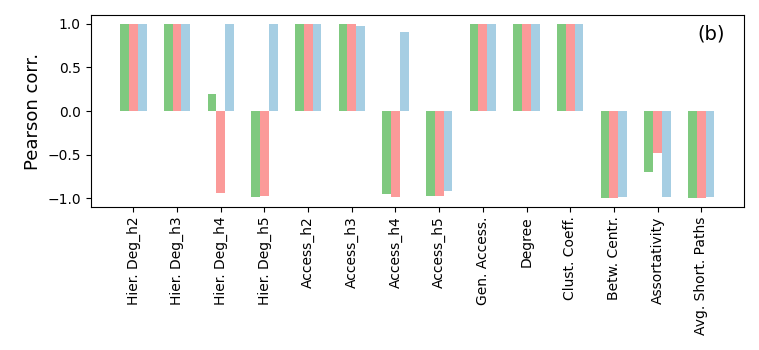}
     \includegraphics[width=0.9 \textwidth]{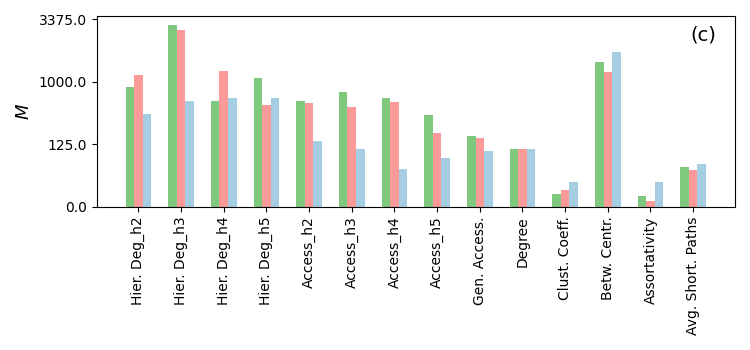}
 \caption{Barplots of the index estimating the overall variation $\psi$ (a); the Pearson correlation (b); and the overall magnitude of the absolute measurement changes (c) obtained for each of the curves in the edge removal experiment. Observe that the vertical axes are presented in non-linear scales in the case of the barplots in (a) and (c) for the sake of enhanced visualization.}\label{fig:barplot_remove}
\end{figure}

Figures~\ref{fig:barplot_remove}(a-c) present, respectively, the overall variation index $\psi$, the Pearson correlation coefficients, and the overall absolute magnitudes of the non-standardized measurement changes obtained for the edge removal experiment.

Interestingly, most the changes undergone by several of the measurements yielded relatively small values of overall variation $\psi$, which can be taken as an indication that these measurement changes are more directly related to the number of edges. Two exceptions can be observed. First, as could be expected, we have virtually null variation in the case of the average degree, as it is directly related to the number of edges. The other exception is that the changes of assortativity have been characterized by a large variation, therefore indicating that this measurement is not specifically associated to the number of edges in the case of the edge removal experiment.

As illustrated in Figure~\ref{fig:barplot_remove}(b), which presents the obtained Pearson correlations, most of the measurement variation curves resulted very close to straight lines, with exception of just a few cases including the hierarchical degree for $h=4$ obtained for ER networks, as well as the assortativity. The above results regarding the overall variation and the Pearson correlation indicate that most measurement changes closely resembled straight lines, while also being specific to the number of edges.

It can be observed from Figure~\ref{fig:barplot_remove}(c) that similar magnitudes of absolute measurement changes have been obtained for the three considered network models. In addition, substantially smaller values have been obtained for the magnitudes of the non-standardized measurement changes of the clustering coefficient and assortativity. At the same time, these two measurement changes are characterized by larger overall variations $\psi$ in Figure~\ref{fig:barplot_remove}(a). 

Of particular interest is the observation that the values of $M$ obtained for the edge removal experiment resulted substantially larger than those obtained in the network size experiment. This is possibly explained by the fact that the progressive removal of edges, while keeping the number of nodes, effectively implies in a variation of the average node degree, which undergoes smaller changes in the case of the network size experiment. This result indicates that the average node degree has marked potential for influencing the properties of the considered networks. 

As shown in Figure~\ref{fig:diagram}, a substantial number of measurements remained into the respective overall groups $A$, $B$, or $C$. However, compared with the previous experiment (network size variations), more measurements changed categories respectively to the network models.  More specifically, the hierarchical node degrees for $h=4$ and $h=5$, as well as the accessibility for $h=5$, belonged to group $A$ in the case of the ER and BA networks but changed to group $B$ in the case of GEO structures. At the same time, the assortativity belonged to group $A$ respectively to the ER and GEO networks but changed to group $C$ in the case of BA structures. As can be seen from Figure~\ref{fig:curves_remove}, the respective $\Delta$ curve is characterized by oscillations as the number of removed edges increases. 

One of the main differences between the results observed in the case of the edge removal experiment and those obtained for the network size experiment consists of the fact that, while most of the measurement changes increased with the number of edges in the case of the latter experiment, now some of the measurement changes decreased with that same free variable. The measurement changes increasing with the average node degree respectively to the three considered network models are: average shortest path, betweenness centrality, assortativity, and access\_h5. 

Recall that, unlike in the case of the network size experiment, where nearby edges are incorporated at once with each respectively added nodes, the average node degree increases with the number of edges, because now the number of nodes is kept constant. This possibly explains why the average shortest path length decreases with the number of edges in the case of the edge removal experiment: a more densely connected network will tend to be characterized by relatively smaller values of shortest path lengths. The fact that the betweenness centrality is defined in terms of the shortest path lengths also possibly explains why this measurement decreased with the average node degree in the case of edge removal. In the case of assortativity, increasing the average node degree as a consequence of the addition of edges in a uniform manner tends to enhance the homogeneity of the network interconnectivity, therefore tending to decrease the type of heterogeneity quantified by the assortativity measurement. The access\_h5 decreasing with the average node degree can be explained by the fact that larger values of the latter property tend to reduce the diameter of the networks, therefore implying that less nodes will be observed at the hierarchical level $h-5$ respectively to each reference node.

\begin{figure}
  \centering
     \includegraphics[width=0.99 \textwidth]{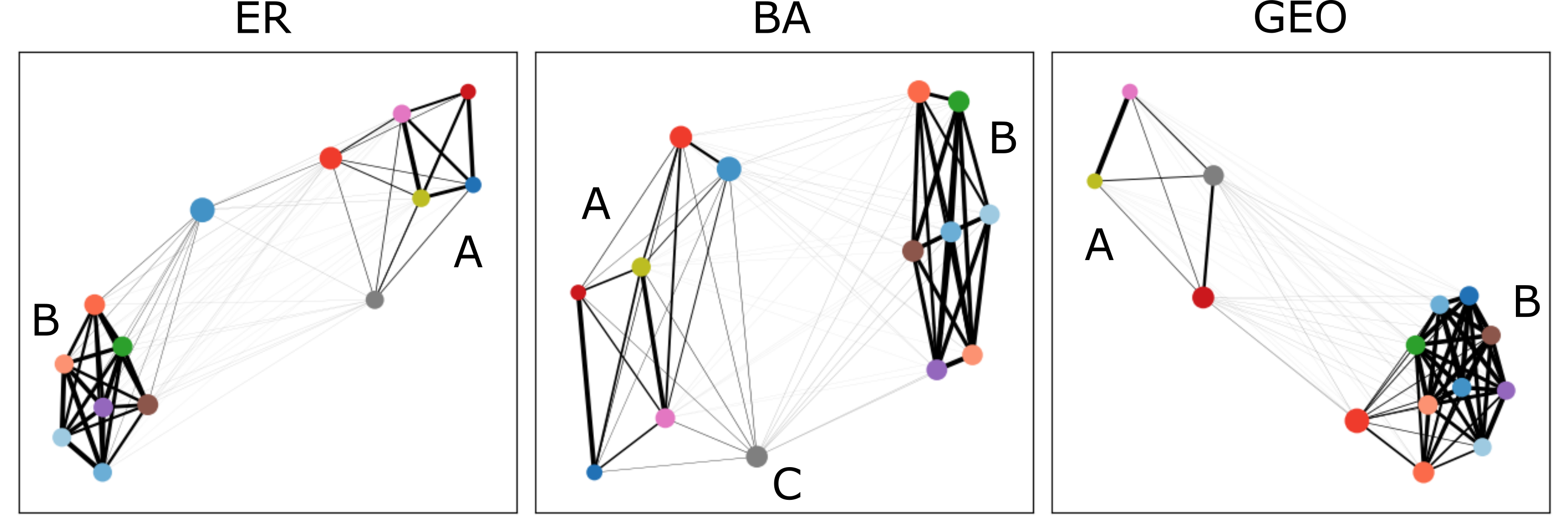}\\
     \hspace{.1cm} (a) \hspace{3.6cm} (b) \hspace{3.6cm} (c)
 \caption{The coincidence similarity networks obtained for the experiments involving removing edges respectively to the following network models: (a) ER, involving three modules; (b) BA, involving two modules; and (c) GEO, involving two components. The color scheme follows the same convention presented in Fig.~\ref{fig:curves_remove}. The width of the edges are proportional to the coincidence similarity index between the respective pair of measurements.}\label{fig:net_remove}
\end{figure}

Figure~\ref{fig:net_remove} illustrates the coincidence similarity networks obtained for the edge removal experiment. Interestingly, these networks can be observe to differ significantly from those obtained for the network size experiment. In particular, two main modules have been obtained for all network models, corresponding to groups of type $A$ and $B$. The BA case includes an additional isolated component corresponding to group $C$. The obtained modularity reflects closely the families of curves observed in Figure~\ref{fig:curves_remove}. At the same time, the networks provide comprehensive additional information about how the curves are interrelated, with a particularly dense cluster of interconnections being observed in the obtained modules of type $B$. Interestingly, the GEO case yielded a similarity network with a module $A$ containing only four measurements.

\begin{figure}
  \centering
     \includegraphics[width=0.99 \textwidth]{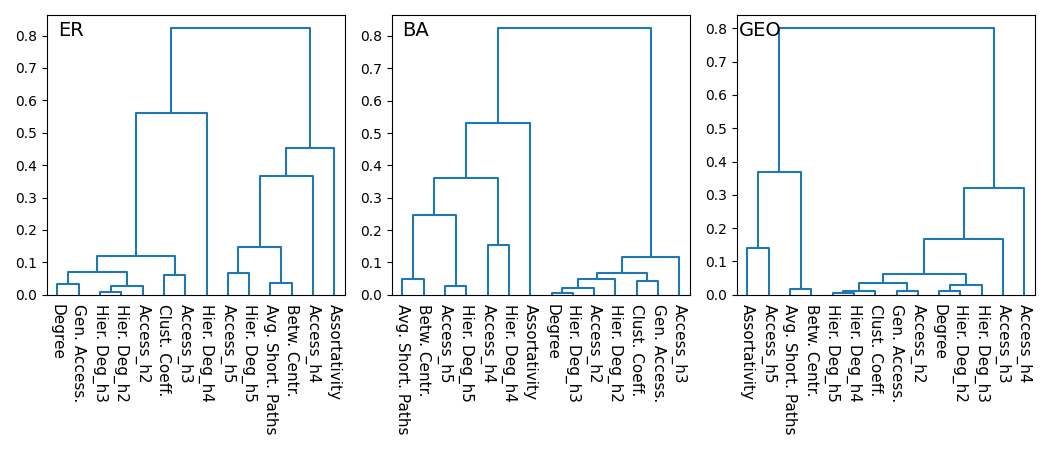}\\
     \hspace{.1cm} (a) \hspace{3.6cm} (b) \hspace{3.6cm} (c)
 \caption{Dendrograms obtained by hierarchical agglomerative clustering of the coincidence similarity networks respectively to the edge removal experiments (Fig.~\ref{fig:net_nodes}). 
 }\label{fig:den_remove}
\end{figure}

The dendrograms obtained for the edge removal experiment are presented in Figure~\ref{fig:den_remove}. Similarly to the coincidence networks, to which they are related, these dendrograms reflect the relationships between the two main groups of curves shown in Figure~\ref{fig:curves_remove}. The particularly compact group of nodes already observed in the case of the respective similarity networks can be readily identified as being associated to the branch with smaller heights in the obtained dendrograms.

\subsection{Edge Rewiring}

This section presents the results and discussion respective to the third experiment involving changes of the topology of ER, BA, and GEO networks, namely concerning respective the progressive rewiring of edges. The number of nodes (network size) and average node degree therefore remains constant along the rewiring.

The considered networks have their size fixed at 100 nodes, all with initial average node degrees are equal to 5.7, and $Q=50$ realizations of the experiment have been performed respectively to each type of network for the number of rewiring edges varying from 0 to 100. 

\begin{figure}
  \centering
     \includegraphics[width=0.99 \textwidth]{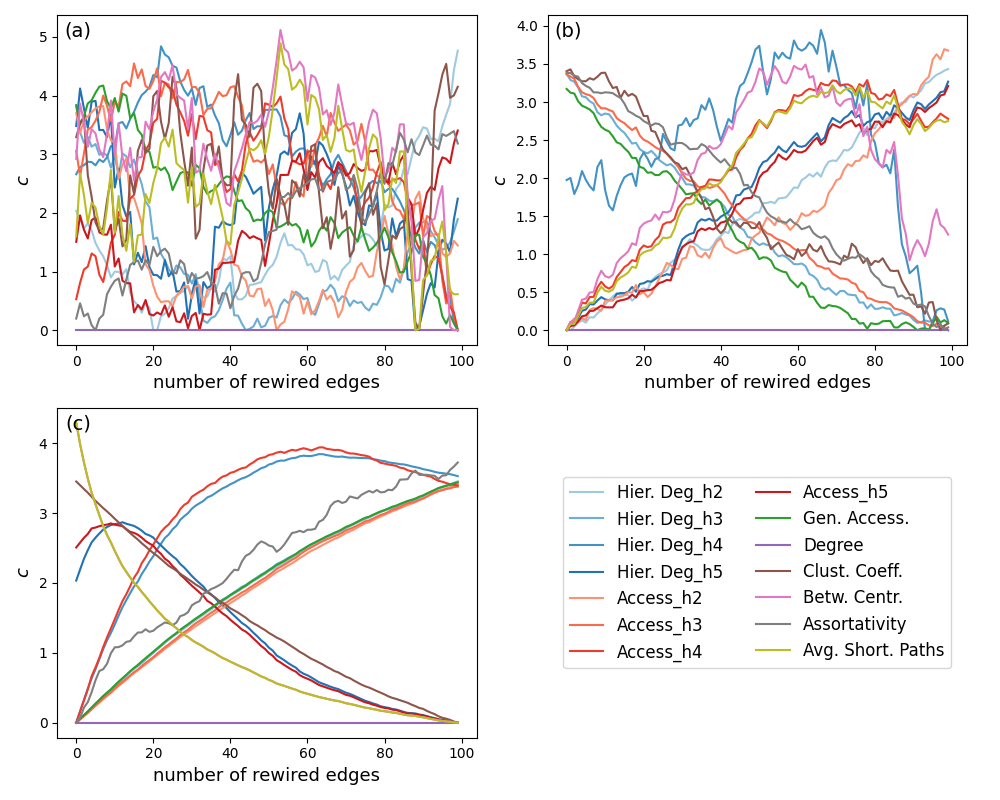}
 \caption{Curves of measurements changes $c$ in terms of the number of progressively rewired edges respectively to the ER (a), BA (b), and GEO (c) types of networks. Markedly distinct results have been obtained for each of these cases.}\label{fig:curves_rewiring}
\end{figure}

Figure~\ref{fig:curves_rewiring} depicts the curves of $c$ in terms of the number of rewired edges respectively to the ER (a), BA (b), and GEO (c) types of networks.

\begin{figure}
  \centering
     \includegraphics[width=0.99 \textwidth]{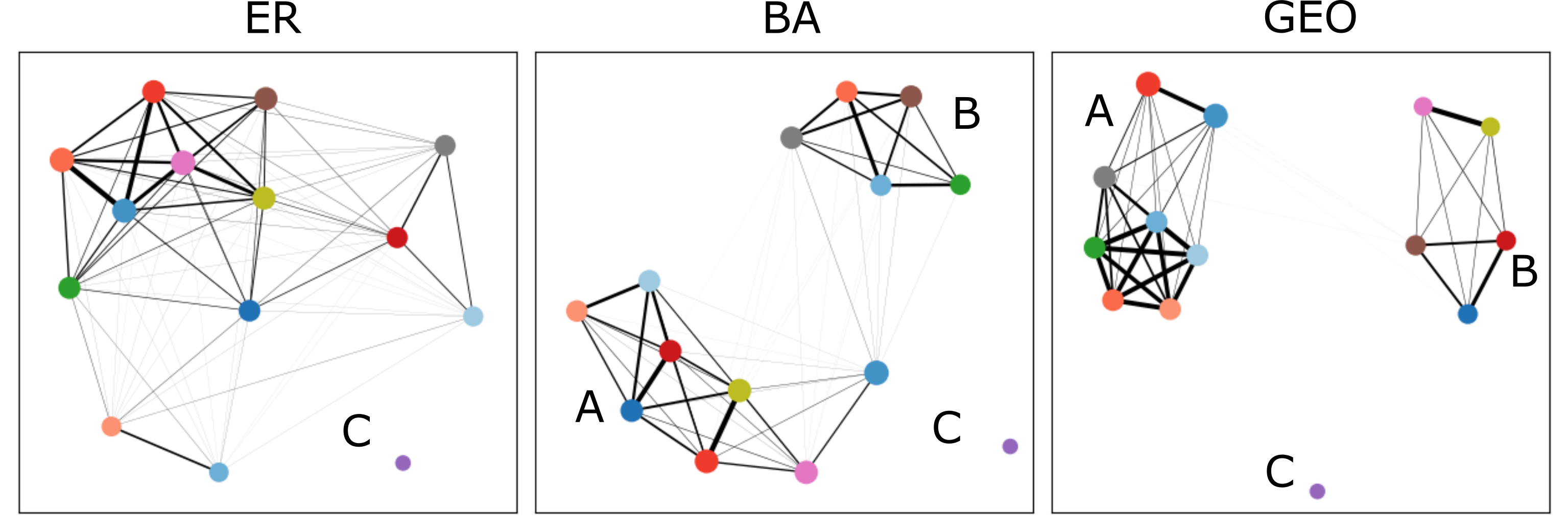}\\
     \hspace{.1cm} (a) \hspace{3.6cm} (b) \hspace{3.6cm} (c)
 \caption{The coincidence similarity networks obtained for the experiments involving the rewiring of the edges respectively to the following network models: (a) ER, involving three modules; (b) BA, involving two modules; and (c) GEO, involving two components. The color scheme follows the same convention presented in Fig.~\ref{fig:curves_rewiring}. The width of the edges are proportional to the coincidence similarity index between the respective pair of measurements.}\label{fig:net_rewiring}
\end{figure}

The $\Delta$ curves obtained for the ER case, except for the case of the average node degree (which remains constant), consist of statistical fluctuations and are shown here only for reference. It should be observed that the relative large variations of these curves in Figure~\ref{fig:curves_rewiring} are a consequence of their respective normalization. This result was expected given that rewiring an ER network yields another network with the same type, size and measurements, except for small statistical variations. On the other hand, the curves of measurements changes $c$ obtained for the BA and GEO networks present moderately defined groups, which are reflected into separated modules in the respective coincidence similarity networks, shown in Figure~\ref{fig:net_rewiring}(b). The membership of each of the measurements variation respectively to the three types $A$, $B$, and $C$ can be seen in Figure~\ref{fig:diagram} diagram. Interestingly, in the case of the GEO networks shown in Figure~\ref{fig:curves_rewiring}(c), most of the measurement variations undergoing predominant decrease also presented a short initial region of relatively small increase.

\begin{figure}
  \centering
     \includegraphics[width=0.9 \textwidth]{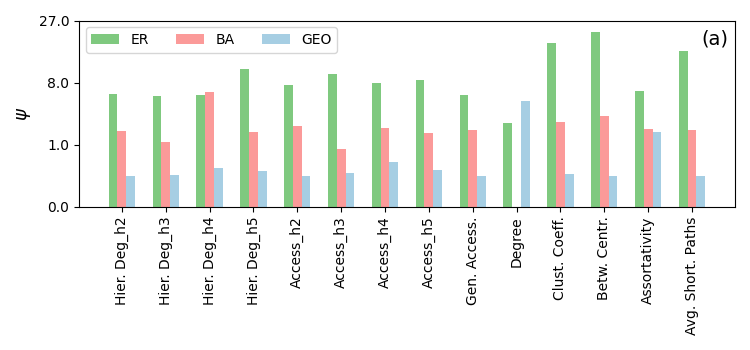}
     \includegraphics[width=0.9 \textwidth]{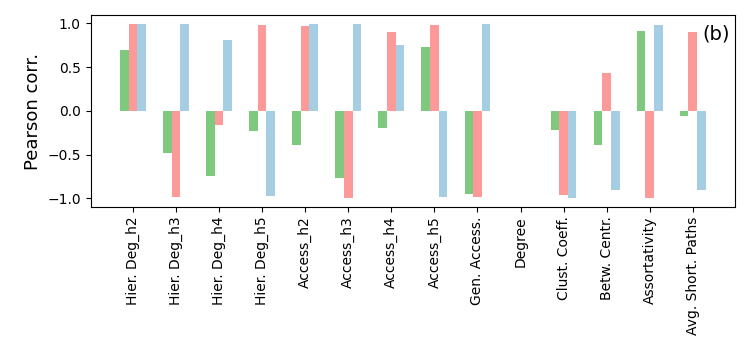}
     \includegraphics[width=0.9 \textwidth]{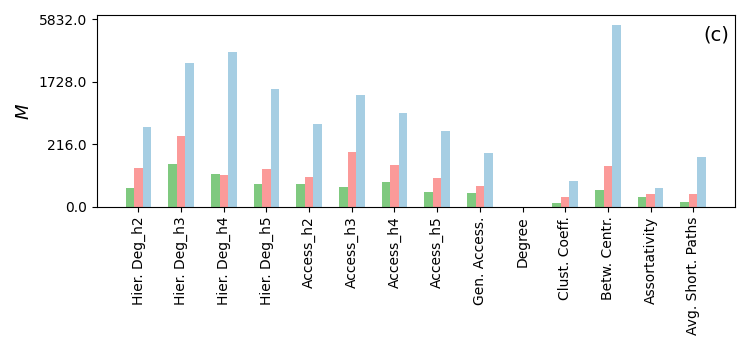}
 \caption{The indices quantifying the overall variation $\psi$ (a), the Pearson correlation (b), and the overall magnitude of non-standardized magnitude changes $M$ (c) obtained for each of the curves in the edge rewiring experiment. Observe that the vertical axes are presented in non-linear scales in the case of the barplots in (a) and (c) for the sake of enhanced visualization.}\label{fig:barplot_rewiring}
\end{figure}

The overall variations $\psi$ and Pearson correlation values obtained for the edge rewiring experiment are presented in Figure~\ref{fig:barplot_rewiring} (a) and (b), respectively. 

In the case of the ER networks (in green), larger values of $\psi$ have been obtained respectively to all measurement change curves, except for the average degree and hierarchical degree for $h=4$. This result stems from the intrinsic structural homogeneity of the ER networks, which implies the measurement changes to correspond to statistical fluctuations. The second and third largest values of $\psi$ were observed for the BA (in red) and GEO (in blue) networks, thus indicating that the latter model of networks has the largest level of stable structural particularities which resulted capable of influencing the measurement changes. 

A markedly distinct set of Pearson correlations, shown in Figure~\ref{fig:barplot_rewiring}(b), have been obtained. These correlations are immaterial in the case of the ER networks, since the measurement changes correspond to statistical fluctuations in that case. Interestingly, relatively large absolute values of Pearson correlation can be observed in the case of GEO networks, indicating that several of the measurement changes take place in a mostly linear manner.

Concerning the overall magnitude of the non-standardized measurement changes $M$, shown in 
Figure~\ref{fig:barplot_rewiring}(c), several results can be observed. First, we have that substantially larger values of $M$ resulted in the case of the GEO networks, while much smaller values followed in the case of the other two network models. Actually, the values of $M$ obtained for the ER and BA cases in the edge rewiring experiment are mostly smaller than the counterparts observed for the two other experiments. This result indicates that edge rewirings have a substantially large effect on the measurements particularly in the case of the GEO networks.

Figure~\ref{fig:net_rewiring} shows the coincidence similarity networks obtained for the edge rewiring experiment. The obtained networks resemble moderately those observed in the previous experiment, namely edge removal.

Given that all measurements changes obtained for the ER networks belong to the category $C$, only the following memberships concerning the BA and GEO networks need to be discussed: (i) $A A$, including \emph{Hier. Deg\_h2}, \emph{Hier.~Deg\_h3}, \emph{Access\_h2}, and \emph{Access\_h4} ; (ii) $A B$, encompassing \emph{Hier.~Deg\_h5}, \emph{Access\_h5}, \emph{Betw.~Centr.}, and \emph{Avg.~Short.~Paths}; (iii) $B A$, incorporating  \emph{Access\_h3},  \emph{Gen.~Access.}, and \emph{Assortativity}; (iv) $C A$, including  \emph{Hier.~Deg\_h4}; (v) $B B$, including  \emph{Clust.~Coeff.}; and (vi) $C C$, encompassing \emph{Degree}.

\begin{figure}
  \centering
     \includegraphics[width=0.99 \textwidth]{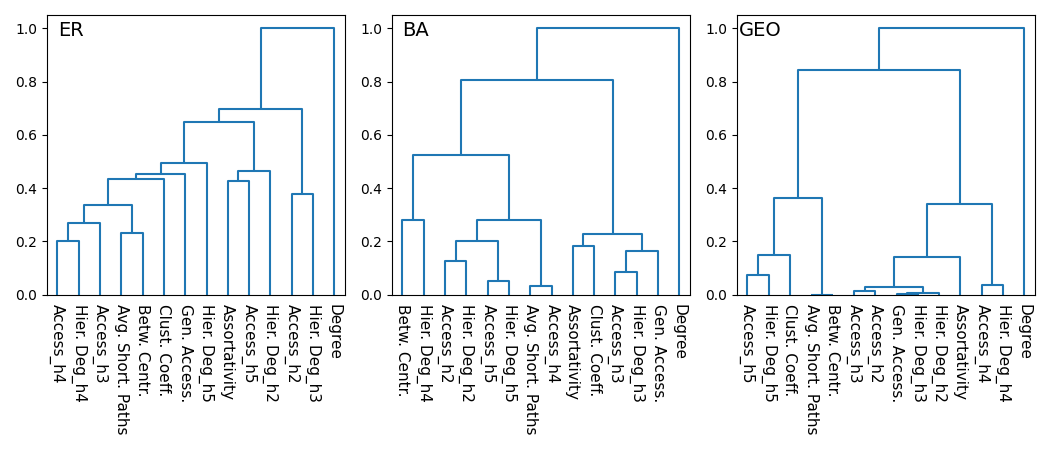}\\
     \hspace{.1cm} (a) \hspace{3.6cm} (b) \hspace{3.6cm} (c)
 \caption{Dendrograms obtained by hierarchical agglomerative clustering of the coincidence similarity networks respectively to the edge rewiring experiments (Fig.~\ref{fig:net_rewiring}).}\label{fig:den_rewiring}
\end{figure}

The dendrograms obtained from the similarity-base hierarchical agglomerative methodology applied on the curves shown in Figure~\ref{fig:den_rewiring} are presented in Figure~\ref{fig:den_rewiring}, respectively to the ER (a), BA (b), and GEO (c) networks.

Interestingly, the fact that the curves obtained for the ER networks consist of statistical fluctuations (Fig.~\ref{fig:net_rewiring}a) has been properly being reflected in closely spaced branches in the respectively obtained dendrogram shown in Figure~\ref{fig:den_rewiring}(a).

\subsection{General Discussion}

The results obtained by employing coincidence similarity networks and similarity-based hierarchical agglomerative clustering respectively to the study of how progressive changes in a network may influence respective topological measurements have been presented and discussed in the preceding subsections respectively to each of three types of progressive changes, namely network size, edge removal, and edge rewiring.

\begin{figure}
  \centering
     \includegraphics[width=0.99 \textwidth]{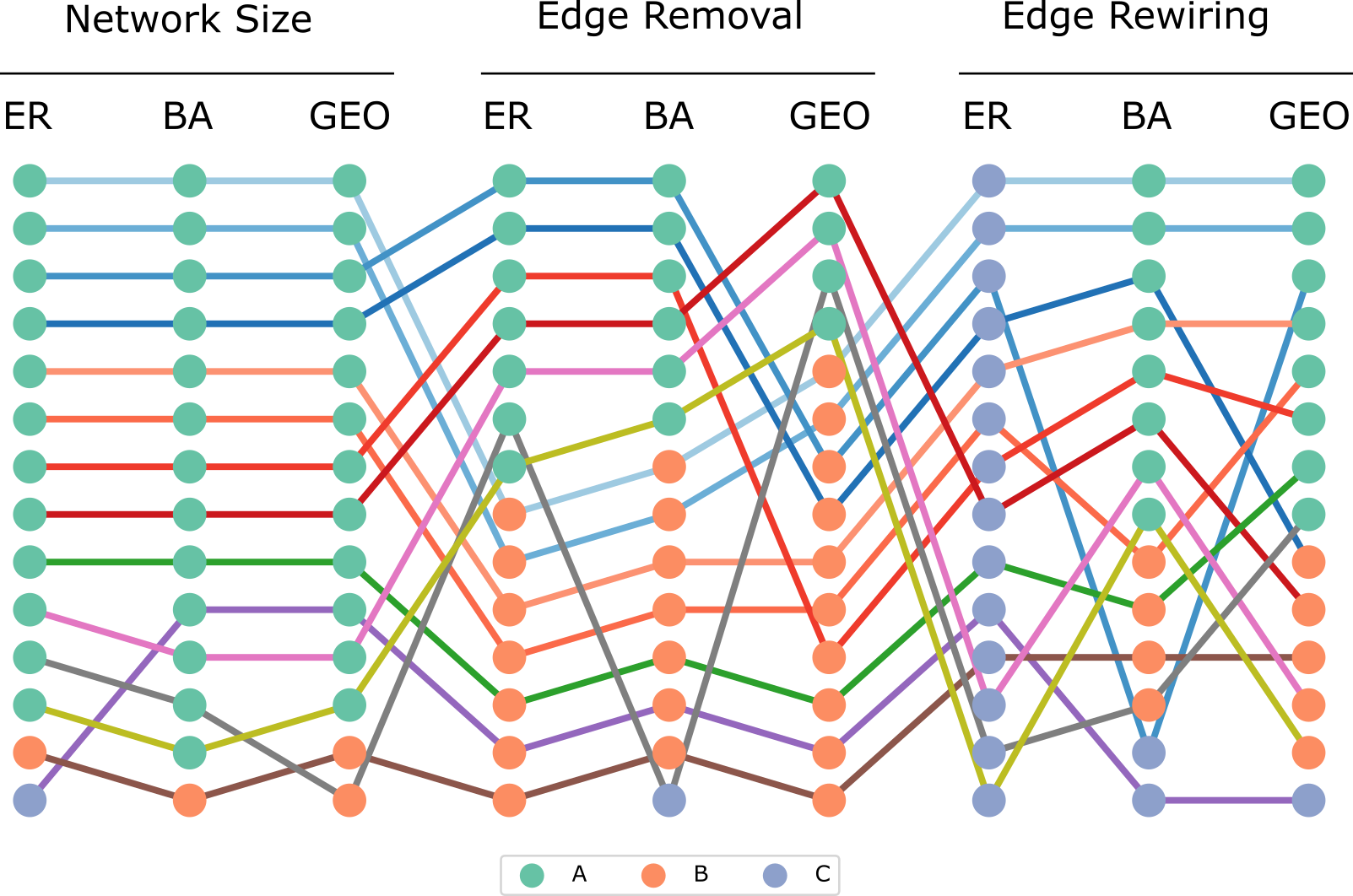}
 \caption{Parallel coordinate presenting the membership of the 14 adopted topological measurements respectively to the categories $A$, $B$, and $C$, identified by three node colors, for the three performed experiments and three models of networks (ER, BA, GEO). The measurements are identified by respective edge colors according to the legend in Fig.~\ref{fig:curves_rewiring}. It can be readily verified that the measurements performed on the ER and BA networks are similarly influenced by the performed network topological changes, but resulted largely distinct for the GEO networks.}\label{fig:diagram2}
\end{figure}

In this section, discussions of the results are presented taking conjointly into account all the three considered experiments. In particular, the classification of the measurements changes into major categories $A$, $B$, and $C$ allowed by the coincidence similarity networks and similarity-based hierarchical agglomerative clustering paved the way to discussing how these groups are preserved or modified respectively to distinct network types and the three considered schemes for progressively changing the networks. Figure~\ref{fig:diagram2} shows the parallel coordinates visualization~\cite{ocagne1885} of the estimated categorization of the measurements changes in terms of the categories $A$, $B$, and $C$.

In the case of the network size experiment, most of the measurement changes have been categorized as belonging to the group $A$, meaning that the changes undergone by these measurements correspond to predominant respective increases with the number of edges (and network size). In addition, this group has been mostly preserved (with the exception of changes implied by the degree and assortativity measurements) among the networks of type ER, BA, and GEO. Thus, we have that most of the measurements in growing versions of these types of networks tended to increase. In the case of the average shortest path length, this effect can be expected, because larger networks with similar average degree will tend to have longer shortest paths. The increase of most of the other measurement is possibly a consequence of the reduction of the border effects which tend to be observed for larger network sizes. Indeed, nodes near to the borders tend to have smaller hierarchical measurements and accessibilities (e.g.~\cite{travenccolo2009border}).

Unlike in the network size experiment, the groups obtained for the edge removals involve a substantial number of measurements being classified into either $A$ or $B$ categories, implying that several measurements categorized as being of type $A$ in the case of the network size experiment changed to the category $B$. More specifically, only the measurements \emph{Access\_h5}, \emph{Betw. Centr.}, and \emph{Avg. Short. Paths} remained in the same category $A$. At the same time, these two major groups tended to remain mostly similar between the ER and BA network types (only the measurement \emph{Assortativity} changed from $A$ to $C$), resulting markedly distinct for the GEO structures, the latter containing 10 measurements of type $B$ and only $4$ of type $A$.

In the case of the edge rewiring experiment in ER networks, the measurements variations were mostly random fluctuations, being therefore categorized into the $C$ group.  However, the coincidence similarity networks obtained for the case of BA and GEO models revealed well-defined groups measurements categorized into the $A$ (8 and 8 cases, respectively), $B$ (4 and 5 cases), and $C$ groups (2 and 1 cases). Interestingly, though having similar number of measurements within each of these groups, extensive changes of measurements memberships can be observed. For instance, in the case of the BA and GEO networks, only the \emph{Hier. Deg\_h2}, \emph{Hier. Deg\_h3}, \emph{Access\_h2}, and \emph{Access\_h4} measurements are shared among the respective groups of type $A$.

\section{Concluding Remarks}

Network science has become an important area in science and technology mainly because of the flexibility of networks in representing and modeling the structural aspects of a wide range of complex systems. Because several types of real-world and theoretical complex networks have been identified, and given that this varying topological structure may impact on respective dynamical properties, substantial effort has been placed on the characterization of their topological characteristics.

In the present work, the effects of three types of changes in three representative complex network types --- namely random (ER), scale free (BA), and geographical( GEO) on a representative set of respective topological measurements have been characterized and studied in terms of representations and visualizations obtained by using coincidence similarity networks~\cite{da2022coincidence} as well as a similarity-based agglomerative clustering methodology~\cite{benatti2024agglomerative}.

The estimation of coincidence similarity networks allowed an effective identification of the main groups of measurements variations, corresponding to modules in the resulting networks. Measurements belonging to a same of these modules tend to present similar response to the imposed progressive topological variations. The reported results and discussions indicate that the obtained network modules tend to correspond to they way in which the measurements increase or decrease with the topological changes, as well as to whether these variations take place according to sublinear, linear, or superlinear manner.

Among the several obtained results -- which are specific to the considered settings, including parametric configurations, types of networks and modifications -- we have that the way in which distinct types of progressive modifications of the networks influence respective topological measurements can depend substantially on both the type of modifications being implemented as well as on the topology of the networks undergoing these modifications. In addition, different types of measurements tended to response distinctly to the modifications. Of particular interest, we have that the most of the measurements modifications in the case of the network size experiment were characterized by respectively progressive increase, a tendency which has been observed for all the three considered network models. The edge removal and edge rewiring experiments yielded a more even separation between measurements which increased or decreased. The similarity networks respectively obtained also indicated that the groups obtained in these two experiments respectively to each network model were approximately similar, involving two main modules corresponding to groups of type $A$ and $B$. However, in the case of the edge rewiring, the changes observed in the case of the ER model mostly corresponded to statistical fluctuations. The obtained dendrograms also contributed for characterizing differences in the intensity of the interrelationship between the responses of the measurements respectively to each considered experiment and network model. The consideration of the bar plots of overall dispersion, Pearson correlation, and magnitudes also helped to complement the interpretation and better understanding of the results of the experiment performed.

Motivated by the intense heterogeneity of the weights obtained in the estimated coincidence similarity networks, a similarity-based hierarchical agglomerative clustering methodology was also adopted as a means to further study the respectively implied interrelationships, providing respective dendrograms. By organizing the measurements variations in a hierarchical manner, the obtained hierarchical analysis indicated that the type of the network can have a major effect in defining the way in which the topological measurements considered change as a consequence of progressive topological modifications. That was the case especially for the GEO networks, which differed more markedly from the results obtained for the ER and BA models, these latter tending to be more similarly affected by the three considered types of progressive network changes. 

All in all, the combined consideration of these approaches allowed an effective complementation of the visualization and analysis of the tendencies of the considered topological measurements to vary as a consequence of progressive topological modifications of the three types of complex networks. Generally speaking, the adopted methodology can be summarized in terms of the three following steps: (i) the curves of the relative alterations undergone by each of the considered measurements are obtained, providing detailed information about the unfolding of the measurements changes as the networks are progressively modified; (ii) the estimation of the coincidence similarity networks is then applied considering the obtained curve in order to reveal in an effective manner how the measurements changes are interrelated, allowing the identification not only of heterogeneous interconnectivity, but also of possible modules, the measurements within each of these modules presenting similar responses to the topological changes; and (iii) the similarity-based hierarchical agglomerative method is then considered as a means of identifying, with the help of respective dendrograms, the hierarchical relationships between the measurements, involving groups and subgroups.
 
The reported concepts, methods, experiments, results, and discussions motivate several further studies. At a more specific level, we have the consideration of other parametric configurations (e.g.~network sizes and average node degree), types and levels of modifications, network models, additional measurements, and types of experiments.  Another promising possibility would be to apply the reported methodology to studying how topological changes of networks may influence respectively implemented dynamics. It would also be interesting to apply the described framework to complement other types of studies.

\section*{Acknowledgments}
Roberto M. Cesar Jr. and A. Benatti are grateful to MCTI PPI-SOFTEX (TIC 13 DOU 01245.010222/2022-44), FAPESP (grant 2022/15304-4), and CNPq. Luciano da F. Costa thanks CNPq (grant no.~313505/2023-3) and FAPESP (grant 2022/15304-4).

\bibliography{ref}
\bibliographystyle{unsrt}

\end{document}